\documentclass{emulateapj}
\usepackage{amsfonts,rotating,lscape}

\newcommand{\si}[1]{\ensuremath{_{\textrm{\tiny{#1}}}}}

\begin{document}

\title{The X-Ray Point-Source Population of NGC\,1365: The Puzzle of Two Highly-Variable Ultraluminous X-ray Sources}
\author{
Iskra V. Strateva\altaffilmark{1},
Stefanie Komossa\altaffilmark{1}
}

\altaffiltext{1}{Max Planck Institute for Extraterrestrial Physics, Postfach 1312, Garching, 85741, Germany}

\begin{abstract}
We present 26 point-sources discovered with Chandra within $200''$ ($\approx20$\,kpc) of the center of the barred supergiant galaxy NGC\,1365. The majority of these sources are high-mass X-ray binaries, containing a neutron star or a black hole accreting from a luminous companion at a sub-Eddington rate. Using repeat Chandra and XMM-Newton as well as optical observations, we discuss in detail the natures of two highly-variable ultraluminous X-ray sources (ULXs): \hbox{NGC\,1365\,X1}, one of the most luminous ULXs known since the \emph{ROSAT} era, which is X-ray variable by a factor of 30, and \hbox{NGC\,1365\,X2}, a newly discovered transient ULX, variable by a factor of $>$90. Their maximum X-ray luminosities  (3--5$\times10^{40}$\,erg\,s$^{-1}$, measured with \emph{Chandra}) and multiwavelength properties suggest the presence of more exotic objects and accretion modes: accretion onto intermediate mass black holes (IMBHs) and beamed/super-Eddington accretion onto solar-mass compact remnants. We argue that these two sources have black-hole masses higher than those of the typical primaries found in X-ray binaries in our Galaxy  (which have masses of \hbox{$<20$\,M$_{\odot}$}), with a likely black-hole mass of \hbox{40--60\,M$_{\odot}$} in the case of \hbox{NGC\,1365\,X1} with a beamed/super-Eddington accretion mode, and a possible IMBH in the case of  \hbox{NGC\,1365\,X2} with  \hbox{$\textrm{M}=80\textrm{--}500\,\textrm{M}_{\odot}$}.

\end{abstract}


\section{Introduction}
Ultraluminous X-ray sources (ULXs) are bright compact off-nuclear point sources detected in nearby galaxies \citep[e.g.,][]{M00,F01,Swartz04}. For sub-Eddington accretion rates in the absence of beaming, their bolometric luminosities of a few$\times10^{39}\textrm{--}10^{40}$\,erg\,s$^{-1}$  require black hole masses $M\si{BH}\sim10\textrm{--}400$\,M$_{\odot}$, implying the existence of intermediate-mass black holes (IMBHs). Despite increasing solid evidence for the existence of supermassive black holes  in sizable bulge galaxies \citep[$M\si{BH}>10^5$\,M$_{\odot}$; e.g.,][and references therein]{Greene}  and the numerous examples of stellar-mass black holes (SMBHs) in our Galaxy and beyond, there are few convincing candidates for black holes with masses in the intermediate regime. The term SMBHs, as used here, refers to black holes produced at the end stage of normal stellar evolution. Observationally, SMBHs are constrained to have masses below $\sim20$\,M$_{\odot}$ \citep[e.g.][and references therein; see also Fryer \& Kalogera~2001 and Orosz et  al.~2007]{RMreview}. Theoretical studies show that compact remnants of up to 60--80\,M$_{\odot}$ could form under special circumstances \citep[assuming solar abundances and initial masses of up to about 200\,M$_{\odot}$; e.g.,][]{H03,Y08} and we adopt this lower limit\footnote{For metallicities much below solar, characteristic of the early universe, black holes with masses over a few hundred solar masses can form \citep{H03}.} as our definition of an IMBH: a compact remnant with mass between 100 and $10^5$\,M$_{\odot}$. The existence of IMBHs presents challenges in terms of formation and feeding mechanisms leading some researches to prefer alternative explanations for ULX emission \citep[e.g.,][]{King,Y08}.  
\begin{deluxetable*}{lcccccc}
\tablewidth{0pt}
\tablecaption{NGC\,1365: \emph{Chandra} X-ray sources}
\tablehead{\colhead{Source} &\colhead{RA} &\colhead{Dec} &\colhead{F\si{0.3-10\,keV}} &\colhead{$\Gamma$} &\colhead{Norm} &\colhead{C-stat}\\ \colhead{(1)} &\colhead{(2)} &\colhead{(3)} &\colhead{(4)} &\colhead{(5)} &\colhead{(6)} &\colhead{(7)}}
\startdata
X-3   & 03:33:21.15 &   -36:08:15.1 &    $1.9_{-1.3}^{+3.6}$ &   1.7$\pm$0.7     &       2.73E-6 &       18/13\\ 
X-4   & 03:33:23.11 &   -36:07:53.1 &   $2.3_{-1.5}^{+2.0}$ &   1.8$\pm$0.3     &       3.71E-6 &       26/25\\ 
X-5   & 03:33:26.03 &   -36:08:38.3 &   $2.2_{-1.8}^{+3.2}$ &   1.5$\pm$0.5     &       2.57E-6 &       17/20\\ 
X-6   & 03:33:26.28 &   -36:08:49.6 &   $2.2_{-1.9}^{+2.2}$ &   1.4$\pm$0.4     &       2.20E-6 &       18/18\\ 
X-7   & 03:33:26.64 &   -36:08:13.9 &   $1.5_{-1.4}^{+1.6}$ &   2 fixed	        &       2.61E-6 &       2/7\\ 
X-8   & 03:33:29.63 &   -36:08:29.9 &   $1.6_{-1.2}^{+1.9}$ &   1.9$\pm$0.5	&       2.77E-6 &       17/19\\ 
X-9   & 03:33:29.64 &   -36:07:56.1 &   $0.8_{-0.8}^{+0.9}$ &   2 fixed	        &       1.48E-6 &       11/8\\ 
X10 & 03:33:30.38 &   -36:08:21.5 &   $0.8_{-0.8}^{+1.0}$ &   2 fixed		&       1.48E-6 &       11/7\\ 
X11 & 03:33:31.09 &   -36:08:07.7 &   $4.5_{-4.5}^{+3.7}$ &   0.1$\pm$0.9     &       6.76E-7 &	    17/10\\ 
X12 & 03:33:31.59 &   -36:08:08.7 &   $5.4_{-2.3}^{+3.0}$ &   1.4$\pm$0.3     &       5.82E-6 &       34/46\\ 
X13 & 03:33:32.22 &   -36:06:42.8 &   $4.6_{-3.6}^{+7.0}$ &   1.1$\pm$0.5     &       3.26E-6 &       16/22\\ 
X14 & 03:33:32.37 &   -36:09:02.9 &   $3.0_{-3.0}^{+10}$ &    0.8$\pm$0.8     &       1.37E-6 &       9/10\\ 
X15 & 03:33:34.16 &   -36:11:02.5 &   $9.8_{-3.0}^{+3.5}$ &   1.5$\pm$0.2     &       1.21E-5 &       61/83\\ 
X1   & 03:33:34.61 &   -36:09:36.6 &   $14_{-3.8}^{+4.3}$ &  1.5$\pm$0.1       &       1.69E-5 &       90/104\\ 
X16 & 03:33:36.45 &   -36:09:58.0 &   $1.0_{-1.0}^{+1.5}$ &   1.1$\pm$0.5     &       7.32E-7 &       6/5\\ 
X17 & 03:33:37.25 &   -36:10:26.2 &   $2.9_{-2.9}^{+7.0}$ &   1.1$\pm$0.7     &       2.14E-6 &       10/10\\ 
X18 & 03:33:38.02 &   -36:09:35.2 &   $8.9_{-3.1}^{+3.3}$ &   1.4$\pm$0.2     &       1.00E-5 &       60/75\\ 
X19 & 03:33:39.21 &   -36:10:01.9 &   $1.3_{-1.3}^{+2.0}$ &   1.5$\pm$0.4     &       1.66E-6 &       14/13\\ 
X20 & 03:33:39.75 &   -36:10:37.3 &   $1.8_{-1.2}^{+1.8}$ &   2.4$\pm$0.5     &       3.80E-6 &       17/21\\ 
X21 & 03:33:40.28 &   -36:07:26.7 &   $4.5_{-2.4}^{+4.2}$ &   1.5$\pm$0.4     &       5.11E-6 &       44/32\\ 
X22 & 03:33:41.94 &   -36:07:41.3 &   $6.0_{-6.0}^{+5.0}$ &  -0.1$\pm$0.5     &       6.32E-7 &	     8/14\\ 
X23 & 03:33:44.48 &   -36:09:58.3 &   $4.0_{-4.0}^{+5.4}$ &   1.2$\pm$0.5     &       3.40E-6 &       15/15\\ 
X24 & 03:33:49.65 &   -36:10:25.6 &   $2.7_{-1.4}^{+1.7}$ &   1.8$\pm$0.3     &       4.53E-6 &       23/31\\ 
\tableline
X25 & 03:33:37.33 & -36:07:23.1 &	$2.4_{-1.9}^{+3.1}$ &	1.1$\pm$0.4	&	1.75E-2 &	22/20\\ 
X2   & 03:33:41.85 & -36:07:31.4 &	$7.8_{-6.2}^{+3.3}$ &	0.5$\pm$0.3	&	2.47E-6 &	47/44\\ 
X26 & 03:33:44.29 & -36:07:17.7 &	$7.7_{-2.8}^{+3.2}$ &	1.2$\pm$0.2	&	6.48E-6 &	75/68\\ 
\enddata
\tablecomments{All but the three sources below the line were detected in our 2002 \emph{Chandra} observation (ObsID~3554); the last three sources are from the 2006 \emph{Chandra} observations with ObsID~6871. Column (1) gives the source designation; (2) and (3) the source position, J2000; (4) the Galactic absorption corrected 0.3--10\,keV flux in units of  $10^{-14}$\,erg\,s$^{-1}$\,cm$^{-2}$; (5) the power-law index of the C-stat fit; (6) the 1\,keV normalization of the C-stat fit in units of photons\,keV$^{-1}$\,cm$^{-2}$\,s$^{-1}$ ; (7) the C-stat per degree of freedom.}
\label{tabSrc}
\end{deluxetable*}

\begin{figure*}
\plottwo{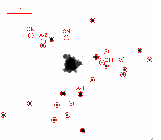}{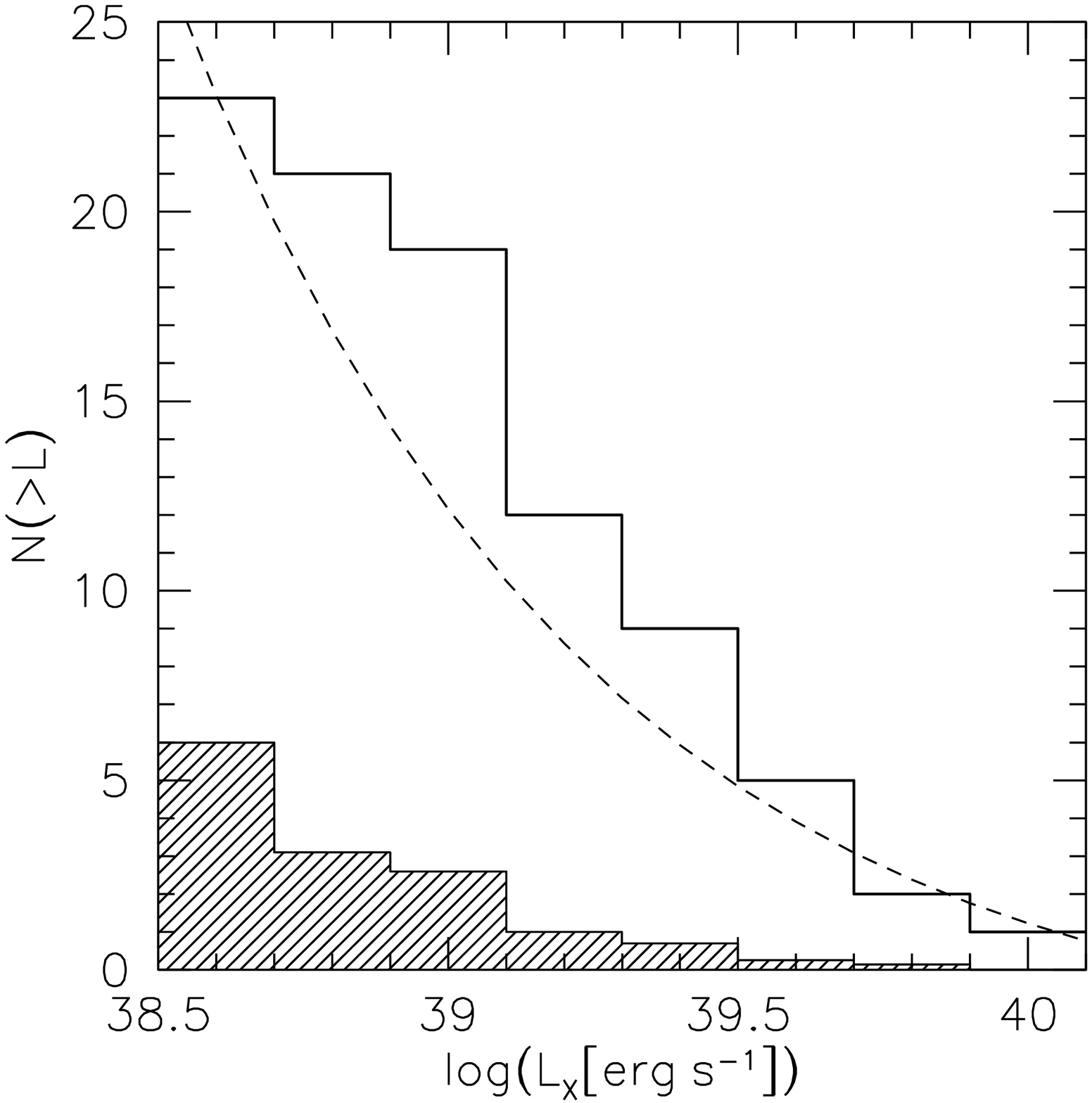}
\caption{\emph{Left:} The background-subtracted, Gaussian-kernel-convolved image of the inner $6'\times6'$ ($\sim36\textrm{\,kpc}\times36\textrm{\,kpc}$) of NGC\,1365 in the 0.3--10\,keV band during our 2002 \emph{Chandra} observation. North is up and East is to the left. The 26 X-ray point sources, detected in \emph{Chandra} observations are denoted by $6''$-radius circles. The two ULXs studied in detail here are marked by X1 and X2. The two point sources marked ``ON''  as well as X2 were much fainter and not detected during the 2002 exposure shown here; the point source marked ``OFF'' fell below the detection limit between 2002 and 2006. \emph{Right:} The cumulative number of X-ray point sources with \hbox{0.3--10\,keV} luminosity greater than a given value for the 23 sources detected during the 2002 exposure (open histogram) and the expected number of background sources based on the  $\log{N}-\log{S}$ results of  \citet{lhole} (hatched histogram). The dashed line is the \citet{GG} prediction for a spiral galaxy with SFR=12\,M$_{\odot}$\,yr$^{-1}$. 
\label{Xgalaxy}}
\end{figure*}

A simple explanation for the observed high X-ray luminosity  of an ULX is a misidentified background active galactic nucleus (AGN), whose chance coincidence with a nearby galaxy and low optical emission contrive to suggest a more unusual object. A recent example is one of the most luminous ULXs found to date in the galaxy MCG-03-34-63 \citep{miniutti}, which was was later identified with a background AGN (Miniutti 2008, private communication). A more intriguing alternative for the origin of the ULX emission than a background AGN is super-Eddington accretion from a high mass X-ray binary (HMXB) with a SMBH primary. If the emission is beamed \citep[e.g.,][]{P07,reljet,G02} and the outflow or jet axis points close to out line of sight, emission of up to $\sim10^{40}$\,erg\,s$^{-1}$ can be explained without the need to invoke IMBHs. Super-Eddington and superluminal accretors have been observed in our Galaxy \citep[e.g. V4641, SS433, GRS\,$1915+105$;][]{V4641,SS433,GRS1905} and according to \citet{King} could be responsible for the apparent super-Eddington emission from the majority of ULXs. Sources with X-ray luminosities in excess of $10^{40}$\,erg\,s$^{-1}$, however, are much rarer and more likely to host IMBHs. Indeed \citet{Swartz04} find that the ULX luminosity function for star-forming galaxies requires a broken power-law fit with break luminosity of $10^{40}$\,erg\,s$^{-1}$,  consistent with the idea of a change in the nature of the ULX population above this luminosity.

\begin{figure*}
\epsscale{1.0}
\plotone{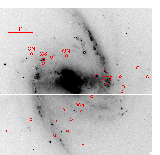}
\caption{VLT FORS2 image of  the inner $6'\times6'$ ($\sim36\textrm{\,kpc}\times36\textrm{\,kpc}$) of NGC\,1365. The 26 X-ray point sources shown in Figure~\ref{Xgalaxy} are denoted by $3''$-radius circles here. North is up and East is to the left. 
\label{Ogalaxy}}
\end{figure*}

Luminous ULXs are more commonly found in star-forming galaxies, leading researchers to suppose that the ULX phenomenon is associated with recent star formation and pointing towards a HMXB origin for the ULX emission. \citet{Rap05} present a detailed analysis of HMXB evolution models together with a binary population synthesis code to study the theoretical expectations for ULX populations of star-forming galaxies under the assumption of a HMXB origin. They conclude that the majority of ULXs with $L\si{X}\lesssim10^{40}$\,erg\,s$^{-1}$ can be explained in terms of binary systems containing SMBHs. Consequently, from both theoretical and empirical perspective, sources with X-ray luminosities in excess of $10^{40}$\,erg\,s$^{-1}$ remain the most puzzling  examples of ULX emission.

In this paper we characterize the X-ray point-source population of the giant barred spiral galaxy NGC\,1365. We show that it is compatible with the population of high mass X-ray binaries expected for galaxies with the star-formation rate of NGC\,1365. \citet{KS98} discussed the first ULX in NGC\,1365, NGC\,1365\,X1, using \emph{ROSAT} HRI data. The source was also noticed by \citet{I97} in an \emph{Advanced Satellite for Cosmology and Astrophysics} (\emph{ASCA}) observation. At the time of discovery this was one of the most luminous ULXs known, which, together with the unusual properties of NGC\,1365 itself, motivated our subsequent 2002 \emph{Chandra} follow-up observation. In this paper, we focus our attention on the X-ray flux and spectral variability of the two brightest ULXs in  NGC\,1365, which reach X-ray luminosities in excess of $10^{40}$\,erg\,s$^{-1}$, and are among the most luminous ULXs known to date \citep[for comparison samples, see][]{M00,F01,Swartz04,D07}.  
Adding optical and infrared (IR) data to the detailed X-ray modeling of \emph{Chandra} and \emph{XMM-Newton} observations, we evaluate the plausibility of the different emission scenarios for the two ULXs. The paper is organized as follows: In \S~\ref{xray}, we present the X-ray point-source population of NGC\,1365. The X-ray spectral and flux variability of NGC\,1365\,X1, together with the available optical/IR data are analyzed in \S~\ref{x1}. We discuss the X-ray and optical data of  the new  ULX, NGC\,1365\,X2, in \S~\ref{x2}, and study in detail the plausibility of the different emission mechanisms for both ULXs in \S~\ref{models}, followed by a short summary in \S~\ref{conclusion}. Throughout this paper we assume a luminosity distance of $D_L=21.2$\,Mpc to NGC\,1365, and the standard concordance cosmology of \citet{Spergel03}. All \emph{Chandra} and \emph{XMM-Newton} data used in this paper were reprocessed using CIAO v.3.4 and HEASOFT v.6.2 (SAS v.7.1) starting with the event 1 files for \emph{Chandra} and repeating the basic pipeline processing for \emph{XMM-Newton} followed by the standard calibration and filtering operations in each case. All quoted flux and luminosity uncertainties are 90\% confidence limits, unless noted otherwise. 

\section{The X-ray Point-Source Population of NGC\,1365}
\label{xray}

NGC\,1365 (RA, Dec = 03:33:36.40, -36:08:25.7; distance modulus\footnote{NASA/IPAC Extragalactic Database, http://nedwww.ipac.caltech.edu/.} of 31.6\,mag) is a luminous giant barred spiral galaxy with a high star formation rate \citep[SFR~$\sim12$\,M$_{\odot}$\,yr$^{-1}$ for $L\si{FIR}=6.8\times10^{10}$\,L$_{\odot}$;][]{L85,K98}. Motivated by the discovery of NGC\,1365\,X1 with \emph{ROSAT}, we observed NGC\,1365 with \emph{Chandra} in 2002,  in order to follow up the luminous ULX and to characterize the X-ray point-source population of the galaxy. The 2002 \emph{Chandra} observation shows diffuse extended nuclear emission and 23 off-nuclear point sources\footnote{Three additional point sources were detected in later \emph{Chandra} observations of the same area; all three are strongly variable and were too faint to be detected during the 2002 \emph{Chandra} observation.} in the central $\sim6'\times6'$ region. The point sources were detected using the \emph{wavdetect} algorithm with source significance $>3\sigma$ \citep{wav} as implemented in CIAO v.3.4. Spectra were extracted for each off-nuclear source from a $2''$-radius aperture and fit with a simple power-law model assuming no intrinsic absorption above the Galactic value using Cash-statistics. The positions, spectroscopic-fit parameters, and model-flux estimates of all off-nuclear point sources are presented in Table~\ref{tabSrc}. Between 2004 and 2006, NGC\,1365 was the target of extensive \emph{XMM-Newton} (P.I. G.~Fabbiano) and \emph{Chandra} (P.I. G.~Risaliti) campaigns aimed at unravelling the highly obscured and variable active galactic nucleus with \emph{Chandra} and studying the ULX NGC\,1365\,X1 with \emph{XMM-Newton}. 

\begin{figure}
\plotone{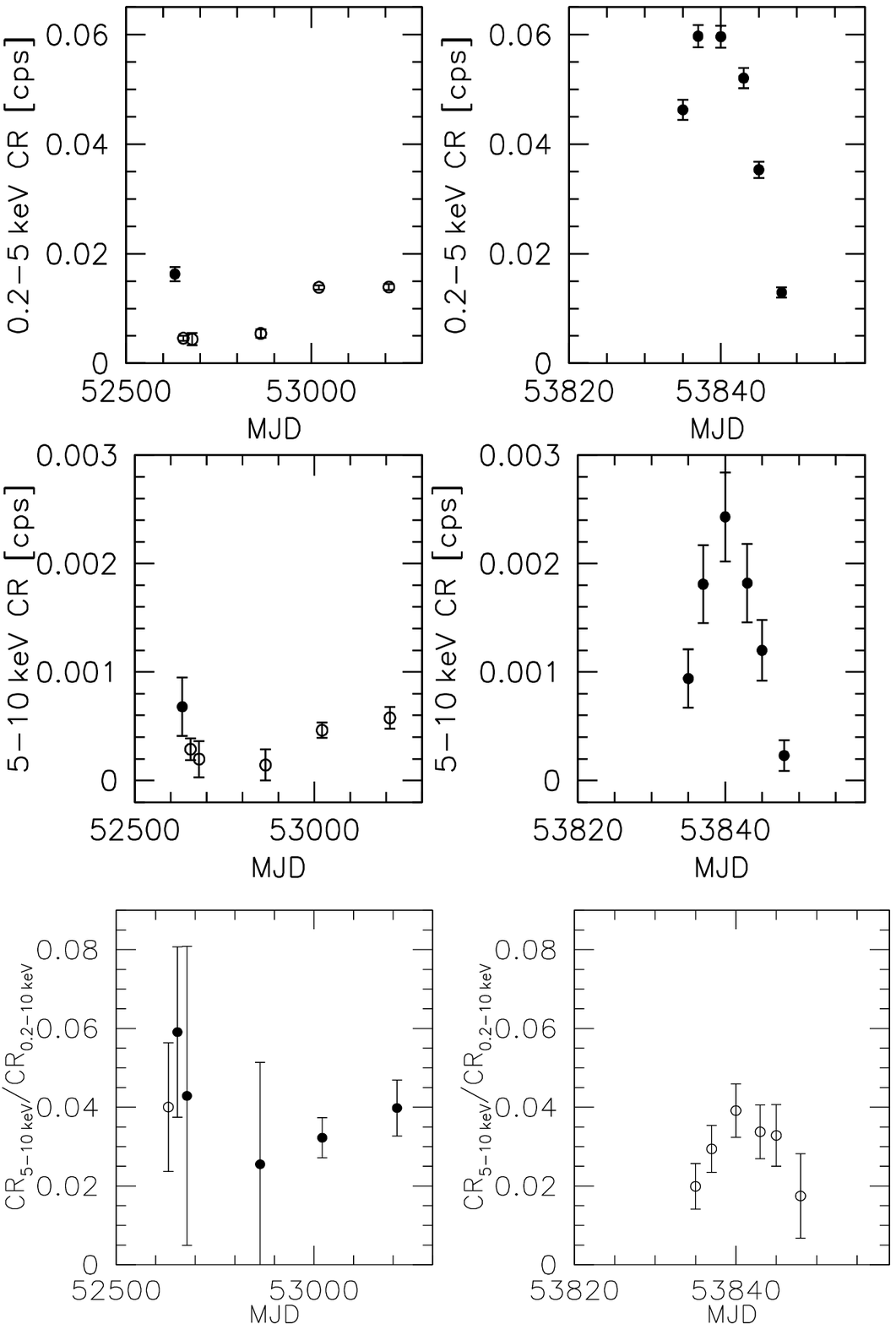}
\caption{NGC\,1365\,X1: light curves in the soft (\hbox{0.2--5\,keV}; top panels) and hard band (\hbox{5--10\,keV}; middle panels) and the fraction of hard (\hbox{5--10\,keV}) vs. total (\hbox{0.2--10\,keV}) emission as a function of time (bottom panels). The solid circles denote the ACIS-S \emph{Chandra} count rates (CRs) while the open  circles denote the ACIS-S CRs equivalent to the observed \emph{XMM-Newton} EPIC-pn CRs. The error-bar propagation assumes Gaussian distribution of the uncertainties and Poisson errors in the measured counts. The right panels show the \emph{Chandra} flare of April 2006.
\label{rates}}
\end{figure}

The left panel of Figure~\ref{Xgalaxy} shows the 2002 \emph{Chandra} background-subtracted and smoothed image of the inner $6'\times6'$ ($\sim36\textrm{\,kpc}\times36 \textrm{\,kpc}$; compare this to the $B=25^m$\,arcsec$^{-2}$ isophotal radius of 30\,kpc). The cumulative number of X-ray point sources with luminosities greater than a given value are shown as a histogram in the right panel. The X-ray luminosities of local star-forming galaxies are proportional to their SFRs \citep[see][and references theirin]{G04} as the numbers of HMXBs, which dominate the extended emission in local star-forming galaxies, increases with increasing SFR. Correcting for the SFR, \citet{GG} have derived a simple analytical expression which predicts the differential number of X-ray point sources N for a given point-source luminosity, $L_{38}$ (the luminosity in units of $10^{38}$\,erg\,s$^{-1}$): $dN/dL_{38}=3.3\times\textrm{SFR}\times L_{38}^{-1.61}$. We plot the integral form of this relation in the right panel of Figure~\ref{Xgalaxy} with a dashed line. The expected number of X-ray point sources in NGC\,1365 agrees well with the observed histogram for bright sources ($\log L\si{X}>38.5$), suggesting that the majority of the X-ray point sources are likely associated with the galaxy. The shaded histogram in the right panel of Figure~\ref{Xgalaxy} is a rough estimate of the possible contamination from background AGNs for each luminosity bin, which we compute using the minimum 2--10\,keV luminosity for each bin  and the $log{N}-\log{S}$ relation of \citet{lhole} as detailed in  \S~\ref{logNlogS}. We expect at most 2 of the 16 sources brighter than $10^{39}$\,erg\,s$^{-1}$ in the $6'\times6'$ area to be background AGNs. We mention in passing that there is no correlation between the measured point-source fluxes and their projected distances from the center of NGC\,1365.

The positions of the 26 X-ray point sources on an optical image of NGC\,1365 are shown in Figure~\ref{Ogalaxy}. Despite the fact that many of the X-ray point sources coincide with the minor and major spiral arms, only 4 out of the 26 have an optical counterpart within $\sim2''$. Three of the four optical counterparts are extended sources, most likely \ion{H}{2} regions, while the remaining southernmost source with an optical counterpart  (X15 in Table~\ref{tabSrc}) is a point source. The optical counterpart (in the case of X2) and upper limits (in the case of X1) are discussed in \S~\ref{x2} and \S~\ref{x1_optical}, respectively. The remaining three sources with optical counterparts, X17, X15, and X13 (see Table~\ref{tabSrc}) have X-ray fluxes of $F\si{0.3--10\,keV}=2.9\times10^{-14}$\,erg\,s$^{-1}$\,cm$^{-2}$, $F\si{0.3--10\,keV}=9.8\times10^{-14}$\,erg\,s$^{-1}$\,cm$^{-2}$, and $F\si{0.3--10\,keV}=4.6\times10^{-14}$\,erg\,s$^{-1}$\,cm$^{-2}$, and power-law spectral indices of $\Gamma=1.1\pm0.6$, $\Gamma=1.5\pm0.2$, and $\Gamma=1.1\pm0.5$ (assuming absorption equal to the Galactic value). The USNO-B\,1.0 \citep{usno} counterparts of X17 and X15 have R-band magnitudes of $R1=18.4$ and $R2=19.2$ (X17) and $R1=19.2$ (X15). 
 \begin{deluxetable}{ccccc}
\small
\tablewidth{0pt}
\tablecaption{NGC\,1365: Observation Summary}
\tablecolumns{5}
\tablehead{\colhead{ObsID} &\colhead{MJD} &\colhead{X1 Counts} &\colhead{X2 Counts} &\colhead{T$_{\textrm{\scriptsize{eff}}}$} \\ \colhead{(1)} &\colhead{(2)} &\colhead{(3)} &\colhead{(4)} &\colhead{(5)}\\ \colhead{\emph{Chandra}} &\colhead{} &\colhead{ACIS-S} &\colhead{ACIS-S} &\colhead{ACIS-S}}
\startdata
3554           & 52632 & 167     & $<4$                &  9.9 \\
6871           & 53835 & 613     & 51                     &  13.4 \\
6872           & 53837 & 880     & 364                   &  14.6 \\
6873           & 53840 & 886     & 306                   &  14.6 \\
6868           & 53843 & 745     & 208                   & 14.6\\
6869           & 53845 & 541     & 117                   & 15.5\\
6870           & 53848 & 190     & 535                   & 14.6\\
\cutinhead{\hskip 0.23in \emph{XMM} \hskip0.35in pn/MOS1/MOS2 \hskip0.8in  pn/MOS}
0151370101   & 52655 & 151/70/84            & ...     &  14.4/17.6 \\ 
0151370201   & 52679 & 24/18/12              & ...     &    2.5/5.4 \\ 
0151370701   & 52864 & 70/15/30              & ...     &    6.6/8.2 \\ 
0205590301   & 53021 & 1519/490/534     & ...     &   49.0/56.5\\ 
0205590401   & 53210 & 999/505/461        & ...    &  30.0/44.6\\
\enddata
\tablecomments{(1) Observation ID; (2) Modified Julian Date (MJD); (3) X1 counts in the  \hbox{0.3--10\,keV} band; (4) X2 counts in the \hbox{0.3--10\,keV} band; (5) The effective exposure time in units of 1000 seconds, after corrections for flaring and chip position.}
\label{tab1}
\end{deluxetable}

\begin{figure}
\epsscale{1.0}
\plotone{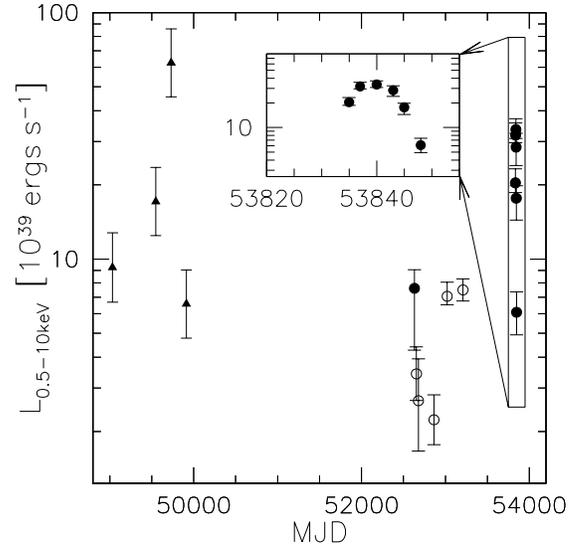}
\caption{NGC\,1365\,X1: luminosity variation with time. The triangles show the \emph{ROSAT}/\emph{ASCA} data, with errorbars indicating only the uncertainty due to the assumed power-law model. The \emph{Chandra} (\emph{XMM-Newton}) power-law model estimated luminosities are shown with solid (open) circles with errorbars indicating the model-fit uncertainties; the inset zooms on the April 2006 \emph{Chandra} flare.
\label{x1lum}}
\end{figure}

\section{NGC\,1365\,X1}
\label{x1}

\begin{figure*}
\plottwo{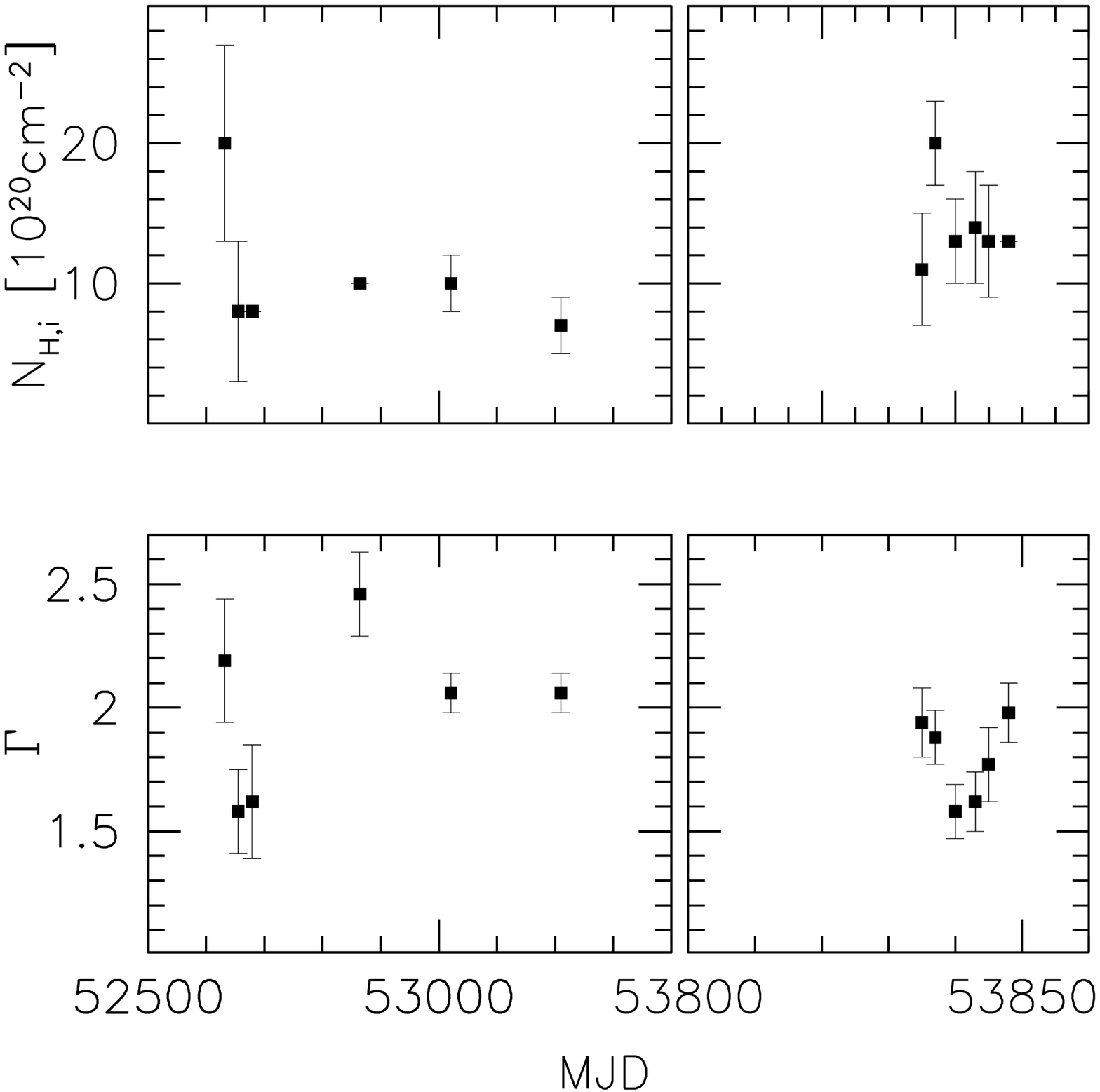}{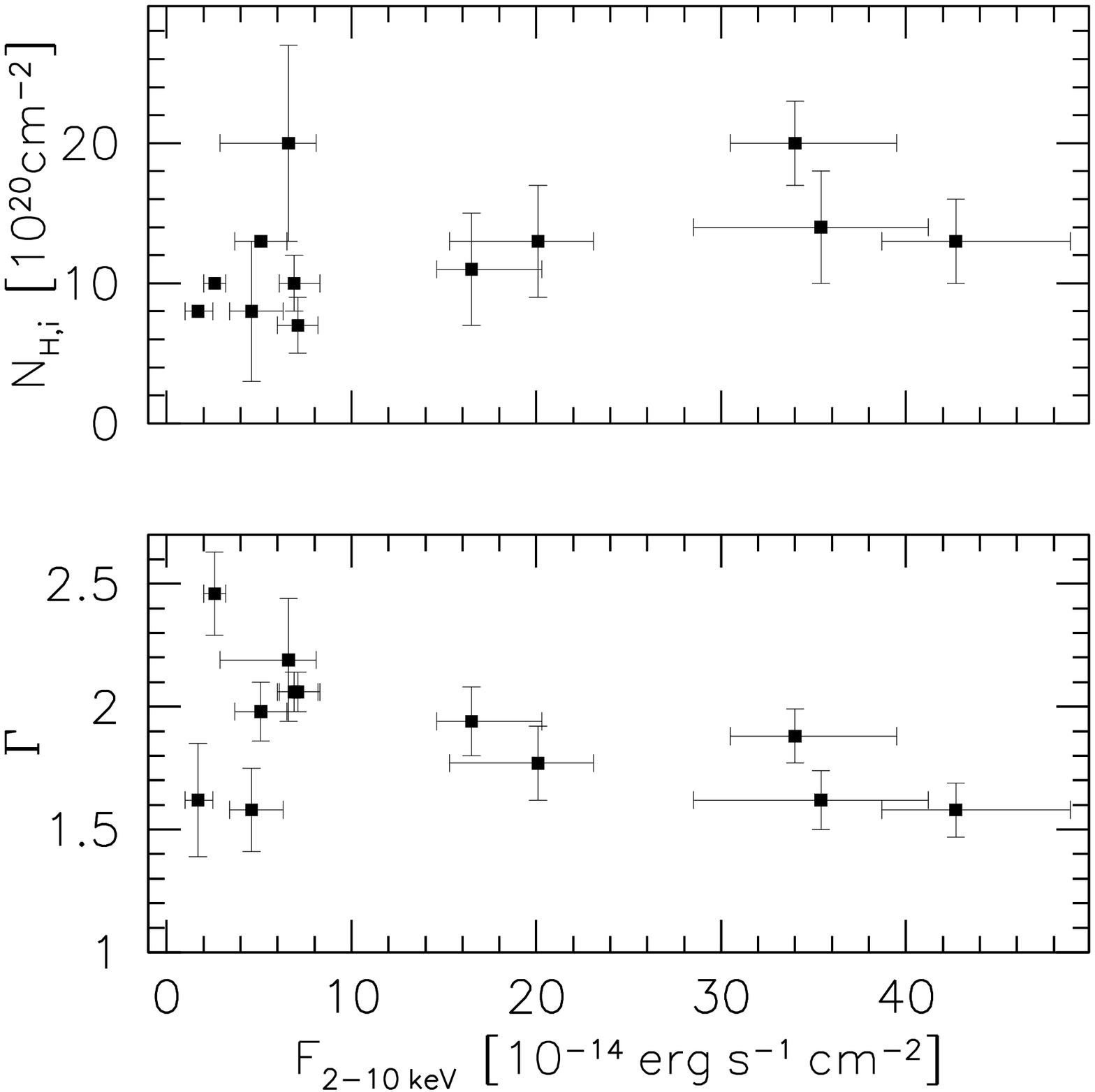}
\caption{NGC\,1365\,X1: variability of the spectral parameters for the absorbed power-law fits with MJD (left panels) and \hbox{2--10\,keV} band flux (right panels).
\label{x1VarPL}}
\end{figure*}

NGC\,1365\,X1 (03:33:34.61, -36:09:36.6) is the most luminous ULX  in NGC\,1365. It was discovered by \citet{KS98} to be highly variable on long timescales (a factor of 10 within 6 months), with a peak luminosity reaching $\sim5\times10^{40}$\,erg\,s$^{-1}$ in the \hbox{2--10\,keV} band \citep[an \emph{ASCA} measurement by][updated to $D_L=21.2$\,Mpc; note that \citet{Soria} find a somewhat lower value by reanalyzing the \emph{ASCA}  spectrum]{I97}.  X1 is $\sim74''$ south from the NGC\,1365 nucleus (equivalent to a projected distance of $\approx7.5$\,kpc) behind an average Galactic column of \hbox{$1.4\times10^{20}$\,cm$^{-2}$}. It is one of the brightest ULXs known, and one of the few ULXs with extensive X-ray (7 \emph{Chandra} and 5 \emph{XMM-Newton} pointings since 2002, in addition to older \emph{ASCA} and \emph{ROSAT} data) as well as extensive optical/IR coverage from the Very Large Telescope (VLT), the Hubble Space Telescope (HST), and the Two Micron All Sky Survey (2MASS). NGC\,1365 was the target of a recent \emph{Chandra} campaign to observe the rapid changes of its obscured active nucleus. Table~\ref{tab1} below gives a summary of these 6 \emph{Chandra} pointings as well as the 2002 \emph{Chandra} and 2003-2004 \emph{XMM-Newton} (5 pointings) observations. We reprocessed all data using CIAO v.3.4.1.1 and HEASOFT v. 6.1.2; the effective exposure times of each pointing, after corrections for flaring and chip positions, are listed in the last column of Table~\ref{tab1}. 

\subsection{X1: X-ray Variability}
\label{xvar}

The soft  (0.2--5\,keV) and hard (5--10\,keV) band light curves of X1 are shown in Figure~\ref{rates}, with the right panels focusing on the recent \emph{Chandra} data. The ULX source underwent a rapid brightening in both soft and hard bands with the soft-band count rate apparently slightly leading the hard band. After varying by less than a factor of $\sim$3 (5) in the soft (hard) band over 2 years, the ULX source flared over the 13-day \emph{Chandra} observation period, changing by a factor of $\sim$5 (10) in the soft (hard) band. Most of the observed counts in both the \emph{Chandra} and \emph{XMM-Newton} exposures are in the soft band below 5\,keV. The bottom panels of Figure~\ref{rates} show the variation of the fraction of hard band photons (using equivalent ACIS-S counts) with time. Note that the source appears to harden during the \emph{Chandra} flare, but as a whole the contribution of the \hbox{5--10\,keV} emission to the total is small, typically 2--4\%.  Despite \emph{XMM-Newton's} higher effective area above 5\,keV ($\sim$ a factor of 3.5 more counts for an EPIC-pn MED filter at the same flux level) there are too few counts in all but the two longest \emph{XMM-Newton} observations above 5\,keV to obtain good constraints on any hard-band components. We return to this point in \S~\ref{xspec} below.

Figure~\ref{x1lum} shows the long term 0.5--10\,keV X-ray variability of NGC\,1365\,X1, including the earlier \emph{ROSAT} and \emph{ASCA} observations.\footnote{We used the ROSAT HRI count rates given in \citet{KS98} and the average power-law model fit obtained in \S~\ref{xspec} below, $\Gamma=1.8$, $N\si{H,i}=10^{21}$\,cm$^{-2}$, which agrees reasonably well with the earlier \emph{ASCA} and PSPC model fits. The errorbars for the \emph{ASCA} and \emph{ROSAT} points indicate only the uncertainties associated with adopting the specific power-law model, ignoring measurement errors and the uncertainties associated with the use of different instruments and passbands.} The X-ray luminosity of NGC\,1365\,X1 is typically a few times $10^{39}$\,ergs\,s$^{-1}$, but varies between $2\times10^{39}$\,ergs\,s$^{-1}$ and $4-6\times10^{40}$\,ergs\,s$^{-1}$, or a factor of $\lesssim 30$ over an eight year period. Note that this large variation on a timescale of years is coupled with much quicker (on the order of days) variation by a factor of 6 during the April 2006 \emph{Chandra} flare. We searched the longest (with non flaring intervals $>10$\,ks) \emph{XMM-Newton} and \emph{Chandra} exposures for variability on scales of minutes to hours but found no statistically significant changes (the probabilities of constant rates were larger than 10\% in all cases). This is consistent with previous results on samples of ULXs, the majority of which show no short term variability \citep[e.g., only 5-15\% of Chandra ULXs are variable on short timescales;][]{Swartz04}.

\begin{deluxetable*}{ccccccc}
\tablewidth{0pt}
\tablecaption{NGC\,1365\,X1: Absorbed Power-Law Fits}
\tablehead{\colhead{ObsID} &\colhead{$F\si{2--10\,keV}$} &\colhead{$F\si{0.5--2\,keV}$} &\colhead{$\Gamma$} &\colhead{N$_{\textrm{\scriptsize{H}}}$} &\colhead{Data/Method} &\colhead{stat/DoF}\\
\colhead{(1)} & \colhead{(2)} & \colhead{(3)} & \colhead{(4)} & \colhead{(5)} & \colhead{(6)} & \colhead{(7)} }
\startdata
3554               & $6.6_{-3.7}^{+1.5}$   & $7.4_{-4.9}^{+2.2}$    & 2.2$\pm$0.2 & 20$\pm$7 & ACIS-S/C-stat     & 80/103   \\
6871               & $16.5_{-1.9}^{+3.8}$ & $21.0_{-2.6}^{+3.6}$ & 1.9$\pm$0.1 & 11$\pm$4 & ACIS-S/$\chi^2$ & 31/34  \\
6872               & $34.0_{-3.5}^{+5.5}$ & $24.6_{-2.1}^{+4.3}$ & 1.9$\pm$0.1 &  20$\pm$3 & ACIS-S/$\chi^2$ & 36/49 \\
6873	               & $42.7_{-4.0}^{+6.2}$ & $19.0_{-3.0}^{+2.0}$ & 1.6$\pm$0.1& 13$\pm$3 & ACIS-S/$\chi^2$ &	50/48 \\
6868               & $35.4_{-6.9}^{+5.8}$ & $17.0_{-4.6}^{+3.4}$ & 1.6$\pm$0.1& 14$\pm$4 & ACIS-S/$\chi^2$ &	36/42 \\
6869               & $20.1_{-4.8}^{+3.0}$ & $12.4_{-3.6}^{+2.7}$ & $1.8\pm0.2$ & 13$\pm$4 & ACIS-S/$\chi^2$ & 18/30\\
6870               & $5.1_{-1.4}^{+1.5 }$ & $6.1_{-1.6}^{+1.9}$ & $2.0\pm0.1$  & 13 fixed & ACIS-S/C-stat  & 101/120\\
0151370101 & $4.6_{-2.0}^{+1.3}$  & $1.7_{-0.7}^{+0.6}$ & 1.6$\pm$0.2   & 8$\pm$5  & PN+MOS/C-stat & 300/332 \\
0151370201 & $1.5_{-0.7}^{+0.8}$  & $3.2_{-1.7}^{+2.2}$ & 1.6$\pm$0.2   & 8 fixed   & PN+MOS/C-stat  & 89/74 \\      
0151370701 & $2.6_{-0.6}^{+0.6}$  & $1.5_{-0.6}^{+0.9}$ & 2.5$\pm$0.2   & 10 fixed    & PN+MOS/C-stat & 151/183 \\
0205590301 & $6.9_{-0.8}^{+1.4}$  & $6.1_{-0.6}^{+1.2}$ & 2.06$\pm$0.08 & 10$\pm$2& PN+MOS/$\chi^2$ & 144/142 \\ 
0205590401 & $7.1_{-1.1}^{+1.1}$  & $6.7_{-0.8}^{+1.0}$ & 2.06$\pm$0.08 & 7$\pm$2  & PN+MOS/$\chi^2$ & 107/107 \\  
\enddata
\tablecomments{The fluxes are quoted in units of $10^{-14}$\,erg\,s$^{-1}$\,cm$^{-2}$. All flux errors are 90\% confidence limits; all other parameter errors are 1$\sigma$ errors. (1) Observation ID; (2) Absorption corrected flux in the 2--10\,keV band; (3) Absorption corrected flux in the 0.5--2\,keV band; (4) Power-law photon index; (5) Intrinsic absorption column density  in units of $10^{20}$\,cm$^{-2}$; (6) Data and fitting method used; (7) $\chi^2/DoF$ or C-stat$/DoF$ of the fit.}
\label{tab2}
\end{deluxetable*}
\begin{deluxetable*}{cccccccc}
\tablewidth{0pt}
\tablecaption{NGC\,1365\,X1: MCD-Power-Law Fits}
\tablehead{\colhead{ObsID} &\colhead{$F\si{diskbb}$} &\colhead{$F\si{0.5--2\,keV}$} &\colhead{$F\si{2--10\,keV}$} &\colhead{$kT$} &\colhead{$K$} &\colhead{$N$} &\colhead{stat/DoF}\\
\colhead{(1)} & \colhead{(2)} & \colhead{(3)} & \colhead{(4)} & \colhead{(5)} & \colhead{(6)} & \colhead{(7)} & \colhead{(8)} }
\startdata
3554               & $4.6_{-1.2}^{+0.3}$ & $4.2_{-2.6}^{+0.2}$ & $9.3_{-1.0}^{+0.3}$ & 0.63$\pm$0.15 & 0.014$\pm$0.011 & 6.5E-6 & 83/106 \\
6871               & $20_{-2.4}^{+1.2}$  & $12_{-2.5}^{+0.5}$  & $19_{-2.6}^{+1.2}$  & 0.86$\pm$0.07 & 0.017$\pm$0.005 & 8.9E-6 & 31/35  \\
6872               & $25_{-2.2}^{+1.0}$  & $14_{-4.1}^{+0.9}$  & $35_{-3.7}^{+1.1}$  & 1.13$\pm$0.10 &  0.0072$\pm$0.026 & 17E-6 & 49/50 \\
6873	               & $23_{-4.0}^{+1.0}$  & $14_{-5.0}^{+0.8}$  & $41_{-7.0}^{+0.9}$  & 1.17$\pm$0.13 & 0.0054$\pm$0.0027& 23E-6 & 49/49 \\
6868               & $22_{-6.2}^{+0.9}$  & $12_{-6.2}^{+1.0}$  & $34_{-9.6}^{+1.1}$  &1.22$\pm$0.13 & 0.0046$\pm$0.023  & 17E-6 & 34/43 \\
6869               & $14_{-3.0}^{+0.9}$  & $8.8_{-3.1}^{+0.5}$ & $20_{-3.1}^{+0.7}$  & 0.97$\pm$0.11& 0.0072$\pm$0.0033 & 11E-6 & 18/31\\
6870               & $6.5_{-3.7}^{+0.5 }$& $3.4_{-2.8}^{+0.5}$ & $5.6_{-2.7}^{+0.5}$ & 0.95$\pm$0.13 & 0.0036$\pm$0.0025 & 2.2E-6 &105/120\\
0151370101 & $2.6_{-1.0}^{+0.3}$ & $1.8_{-1.1}^{+0.3}$ & $4.6_{-1.1}^{+0.3}$ & 1.02$\pm$0.20 & 0.0012$\pm$0.0014 & 2.7E-6 &297/333 \\
0151370201 & $1.8_{-0.7}^{+0.3}$ & $1.2_{-0.8}^{+0.3}$ & $3.3_{-0.7}^{+0.3}$ & 1.04$\pm$0.52 & 0.0007$\pm$0.0023 & 1.9E-6 & 88/74 \\      
0151370701 & $2.8_{-1.0}^{+0.3}$ & $2.0_{-1.4}^{+0.2}$ & $2.0_{-0.3}^{+0.2}$ & 0.46$\pm$0.07 & 0.028$\pm$0.017& 1.3E-6& 151/183 \\
0205590301 & $6.0_{-1.2}^{+1.0}$ & $4.5_{-0.3}^{+0.2}$ & $7.3_{-2.1}^{+1.9}$ & 0.65$\pm$0.04 & 0.015$\pm$0.003& 4.4E-6 & 151/142 \\ 
0205590401 & $6.7_{-0.6}^{+0.5}$ & $5.3_{-0.4}^{+0.2}$ & $8.1_{-0.3}^{+0.2}$ & 0.52$\pm$0.03 & 0.041$\pm$0.008 & 5.5E-6 & 99/108 \\  
\enddata
\tablecomments{The fluxes are quoted in units of $10^{-14}$\,erg\,s$^{-1}$\,cm$^{-2}$. (1) Observation ID; (2) Flux of the MCD component in the 0.0001--10\,keV band;  (3) Total flux in the 2--10\,keV band; (4) Total flux in the 0.5--2\,keV band; (5) Maximum MCD-fit  temperature in keV;  (6) Normalization of the MCD component, $K=(R\si{in}/D)^2 \cos{\theta}$, where $R\si{in}$ is the apparent inner accretion-disk radius in km, $D$ is the distance to the source in units of 10\,kpc and $\theta$ is the inclination of the disk; see \citet{Kubota}; (7) 1\,keV normalization of the $\Gamma=1$ power-law component in units of photons keV$^{-1}$\,cm$^{-2}$\,s$^{-1}$; (8) $\chi^2/DoF$ or C-stat$/DoF$ of the fit; the type of the fit for each observation ($\chi^2$ or C-stat) is the same as that given in Table~\ref{tab2}. We assume no intrinsic absorption above the Galactic value of $1.4\times10^{20}$\,cm$^{-2}$ for all fits. All flux errors are 90\% confidence limits; all other parameter errors are 1$\sigma$ errors. }
\label{tab3}
\end{deluxetable*}

\subsection{X1: Spectroscopic Fits}
\label{xspec}

\begin{figure*}
\plotone{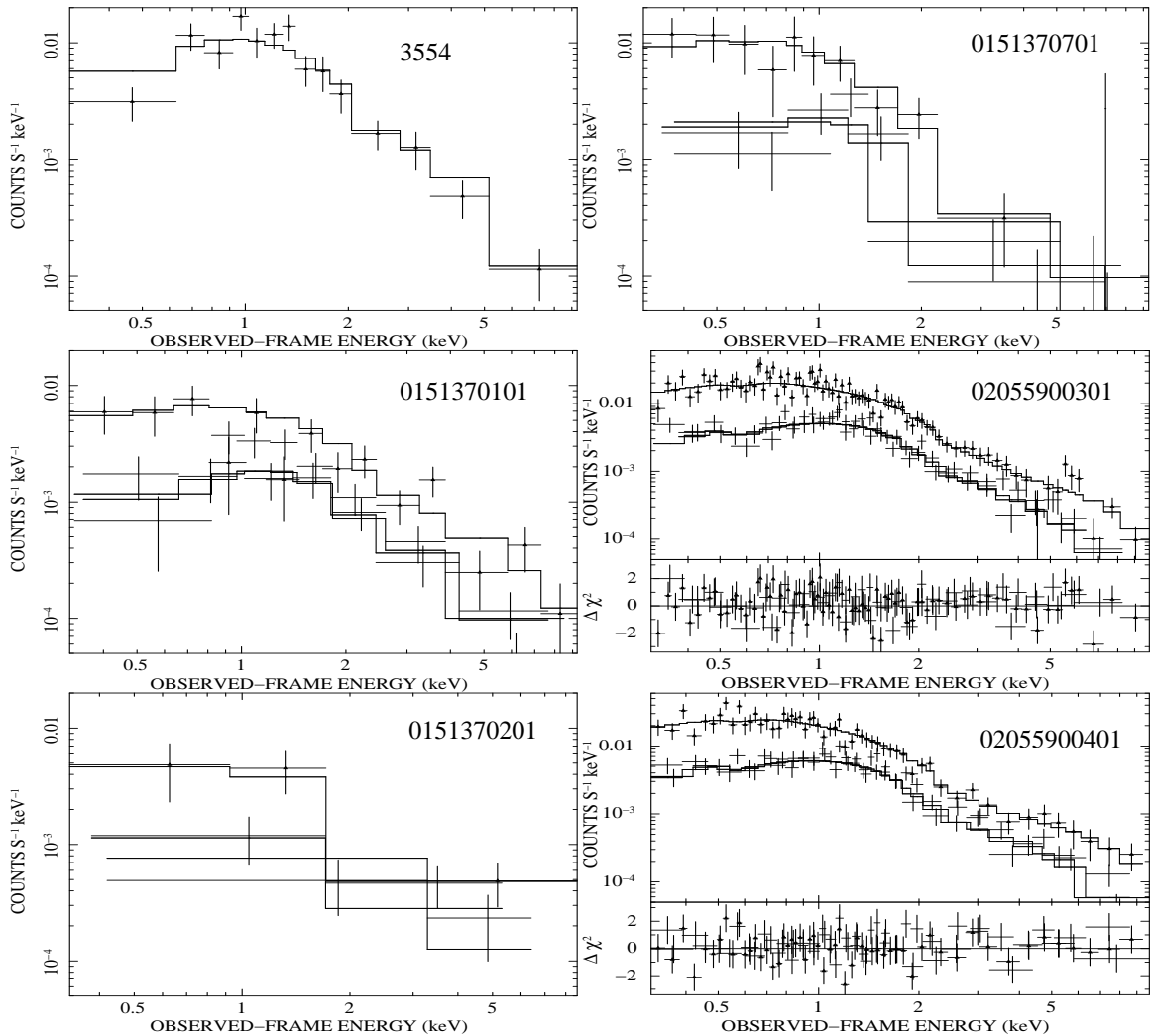}
\caption{NGC\,1365\,X1: MCD+power law fits to the 0.3--10\,keV \emph{Chandra} and \emph{XMM-Newton} spectra. The MJD increases from top to bottom and left to right. The ObsIDs are given in the top right corner of each panel. The apparent residuals around \hbox{5--6\,keV} in \emph{XMM-Newton} observation 0205590301 are likely an artifact of the background subtraction, since in this case the obscured active nucleus was particularly bright relative to X1 in the hard band and this ``feature" is absent from spectra extracted from regions with smaller extraction radii.
\label{x1spec1}}
\end{figure*}

\begin{figure*}
\plotone{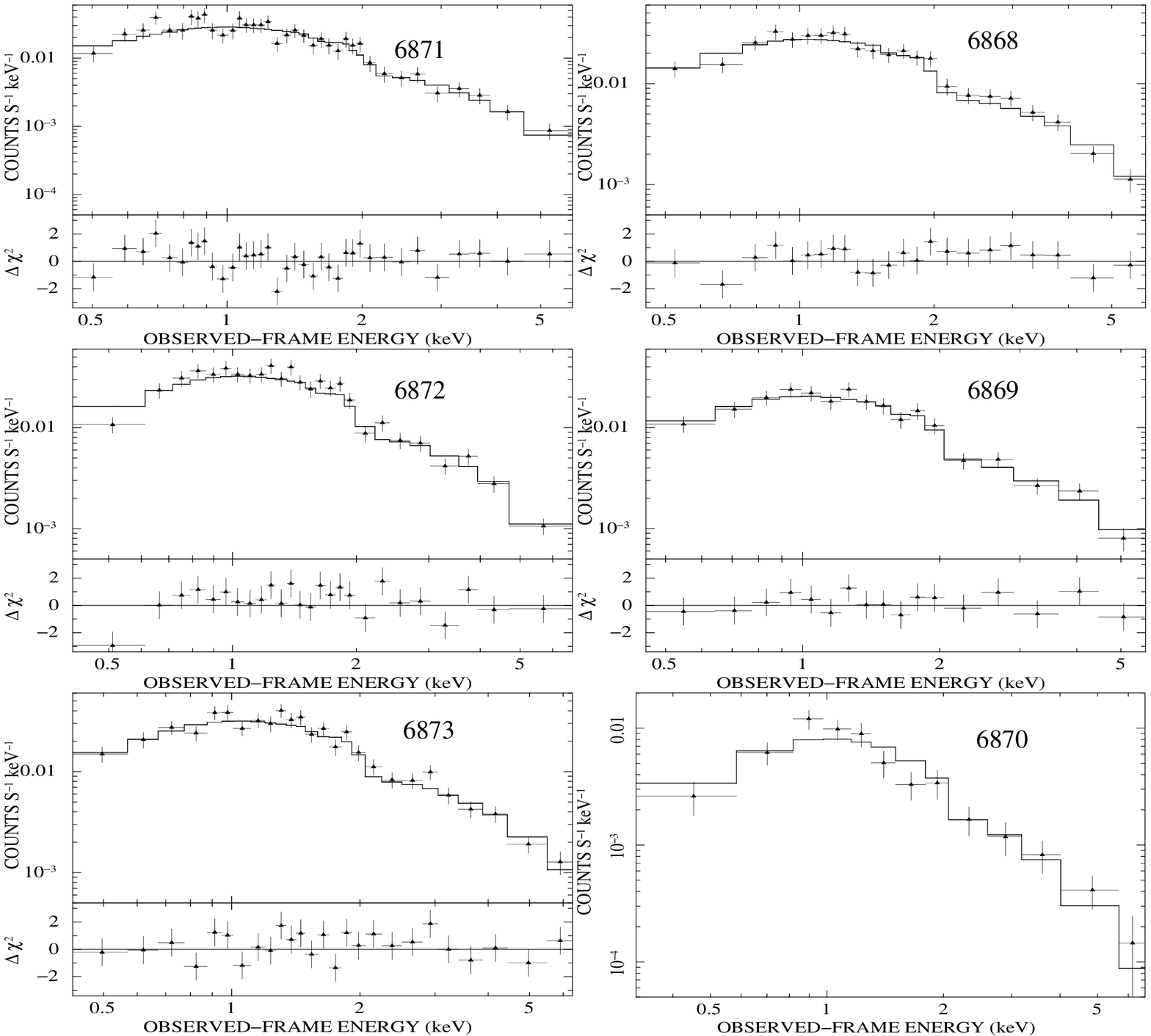}
\caption{NGC\,1365\,X1: MCD+power-law fits to the 0.3--10\,keV April 2006 \emph{Chandra} spectra. The MJD increases from top to bottom and left to right. The ObsIDs are given in the top right corner of each panel. 
\label{x1spec2}}
\end{figure*}

We extracted spectra from $2''$-radius (\emph{Chandra}) or $15''$-radius (\emph{XMM-Newton}) regions\footnote{The size of the \emph{Chandra} extraction region corresponds to over 90\% encircled energy. For \emph{XMM-Newton} the $15''$-radius extraction region was selected to minimize contamination from the obscured active nucleus in the case of X1 and the encircled energy is expected to be about 65--70\%. We extracted spectra from larger radii (up to $45''$) for the two longest  \emph{XMM-Newton} exposures and found that the flux estimates using these larger radii are consistent with the ones quoted, within the errorbars.} around NGC\,1365\,X1 and selected background regions free of point sources at similar distance from the NGC\,1365 nucleus to minimize contamination. The spectral fits were performed using Cash-statistics for individual spectra with less than 300 counts (two \emph{Chandra} and three \emph{XMM-Newton} observations marked ``C-stat" in Table~\ref{tab2}) using 15-count binning for the remaining five \emph{Chandra} and 20-count binning for the two longest \emph{XMM-Newton} observations. In order to test specific models of X-ray emission we attempted to fit all spectra with two simple models: (1) power-law fits, including intrinsic absorption; (2) multi-color disk (MCD) plus power-law emission models with absorption equal to the Galactic value. The second model is physically motivated, representing the optically thick thermal X-ray emission from the accretion disk (\emph{diskbb} model in XSPEC) as well as a harder, corona comptonized component (represented by a simple \emph{powerlaw} model in XSPEC). The spectral fits for each of the \emph{Chandra} and \emph{XMM-Newton} observations for the two different models are given in Tables~\ref{tab2} and \ref{tab3}, respectively. 

\begin{figure*}
\plottwo{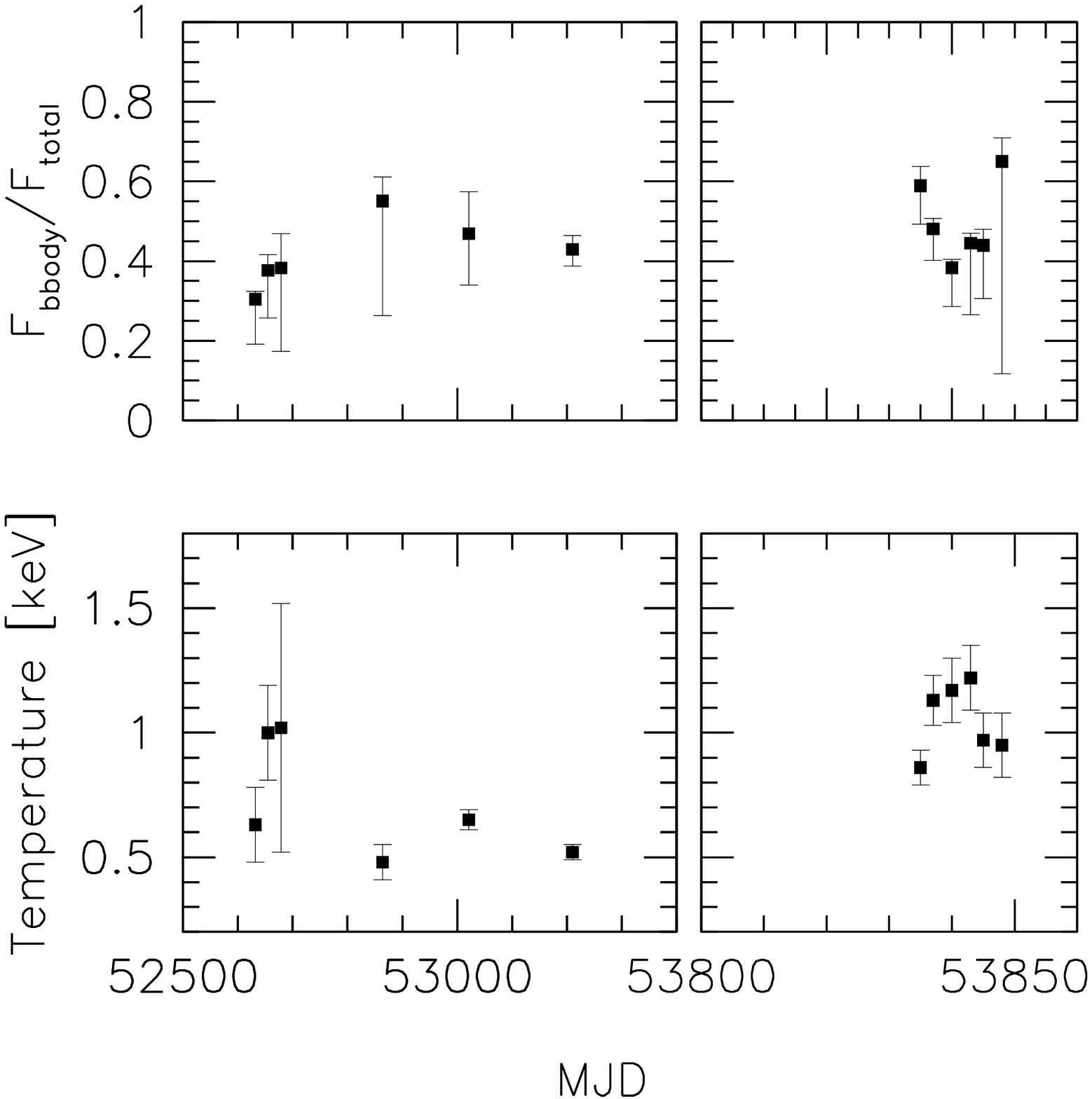}{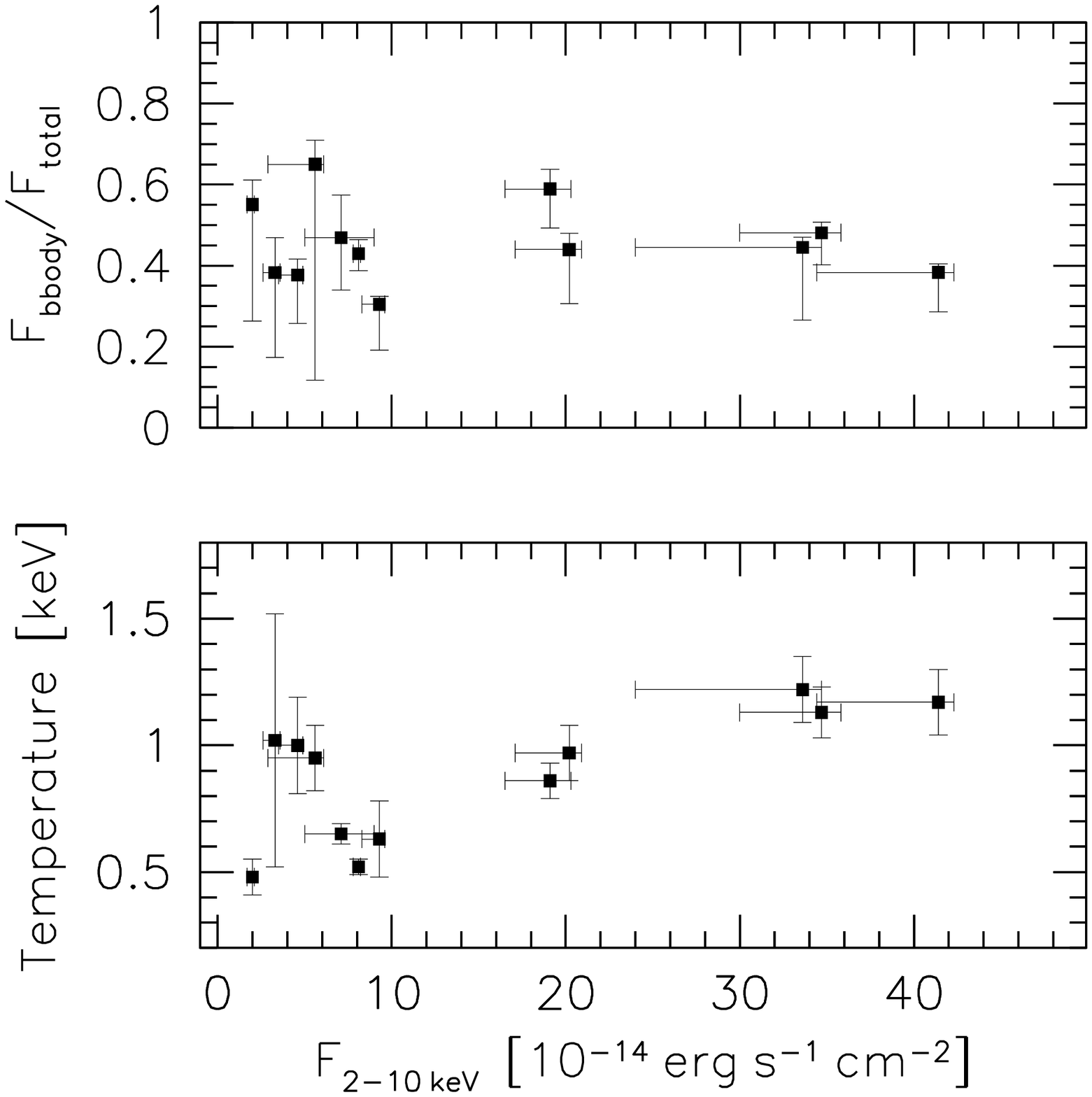}
\caption{NGC\,1365\,X1: variability of the spectral fit parameters for the MCD+power-law fits with MJD (left panels) and \hbox{2--10\,keV} band flux (right panels).
\label{x1VarMCD}}
\end{figure*}

The addition of an intrinsic absorber substantially improved the power-law model fits for 9 of the 12 observations shown in Table~\ref{tab2}. For the remaining three observations with limited photon statistics, we fix the intrinsic absorption to the value estimated for the well constrained observation closest in time.  The variability of the spectral parameters for the power-law fits are shown in Figure~\ref{x1VarPL}. The required intrinsic absorption, $N\si{H,i}$, ranges between ($7\pm2$) and $(21\pm3) \times10^{20}$\,cm$^{-2}$, a variation by a factor of $\lesssim3$. Similar increase in the absorbing column with increasing flux was observed in the HMXB Cygnus-X3 with INTEGRAL \citep{CygX3}. The estimated power-law photon indices vary between $1.6\pm0.1$ and $2.5\pm0.2$, with the majority of fits having $\Gamma=1.9\textrm{--}2$. From Figure~\ref{x1VarPL}, the spectral index appears to flatten with increasing flux during the \emph{Chandra} flare of April 2006 (indicating a harder spectrum) while the intrinsic absorption appears to remain constant over the same 13-day period. Note that the spectral parameter variation in general needs to be interpreted with caution, due to the known degeneracy between $\Gamma$ and $N\si{H,i}$ in power-law fits, as well as the lower sensitivity of \emph{Chandra} at high energies \citep[see also the discussion in Appendix A3 of][]{imbhulx1}. 

Figures~\ref{x1spec1} and \ref{x1spec2} show the \emph{Chandra} and \emph{XMM-Newton} fits for the second model, representing the direct accretion disk emission plus corona-comptonized component. Statistically these fits are indistinguishable from the absorbed power-law fits presented above (compare the last columns of Tables~\ref{tab2} and \ref{tab3}). In fact if we only consider the \hbox{0.3--5\,keV} emission, the MCD fits are adequate and no power-law component is necessary in any of the 12 observations. Considering the full \hbox{0.3-10\,keV} band, the MCD model alone provides good fits to all but the three longest \emph{XMM-Newton} observations, for which the addition of a power-law component  improves the fit with $>$99\% confidence. As noted in \S~\ref{xvar} the \emph{Chandra} and  (to a lesser extent) \emph{XMM-Newton} effective areas result in insufficient counts above 5\,keV for all but the two longest \emph{XMM-Newton} exposures. The additional power-law component required by these spectra has a  photon index of $\Gamma\approx1$. Fitting the longest exposure, \emph{XMM-Newton} 0205590301, we obtain the normalization of this additional power law component and proceed to include a   $\Gamma=1$ power-law component in all MCD fits, scaling the normalization of the power law according to the observed \hbox{5--10\,keV} count rate.\footnote{Using \emph{XMM-Newton} observation 0205590401 instead of 0205590301 to obtain the power-law photon index and normalization does not affect the results.} The resulting MCD plus power-law fit parameters are listed in Table~\ref{tab3} and their variability with time and X-ray flux is shown in Figure~\ref{x1VarMCD}. These fits allow us to estimate both the maximum color temperature of the accretion disk and the total luminosity emitted by the disk and to compare these to IMBH and supercritically accreting SMBH models in \S~\ref{LT}. In this model, the accretion-disk emission contributes 30--60\% of the total flux in the 0.3--10\,keV band and the maximum color temperature increases with increasing flux. 

\citet{Stobbart}, who studied the highest S/N \emph{XMM-Newton} spectra of 13 ULXs, concluded that neither simple power law models nor simple MCD models provide adequate fits to the highest S/N spectra of ULXs. Better representations of their data are found using cool ($\sim$0.2\,keV) or warm ($\sim$1.7\,keV) disk plus power-law models, combination of cool black body plus warm MCD (``dual thermal") models (\emph{bbody}$+$\emph{diskpn} in XSPEC), or MCD plus comptonized-corona models  (\emph{diskpn$+$eqpair} in XSPEC). The authors caution that depending on the choice of model, the inferred disk temperatures of their sample would cluster either at the ``cool" or ``warm" end of a luminosity-temperature diagram, and be interpreted as evidence of the presence of IMBHs (if cool) or SMBHs (warm). Even the best spectra (e.g., over $10^4$ counts with \emph{XMM-Newton}) were inadequate to distinguish between these models. Thus, disk-temperature measurements based on spectra with less than a few thousand counts could give misleading results, especially when using the ``wrong" spectral model. In addition, \citet{Stobbart} suggest that the evidence for a distinct curvature in the 2--10\,keV spectra of ULXs (seen as a break in the power-law emission in 8/13 of their sources) can be used as a red flag against the interpretation of the hard band spectrum as emission from an optically thin corona. A broken power law model above 2\,keV is not required by any of our individual observations nor by the joint fit to the highest S/N \emph{XMM-Newton} and/or \emph{Chandra} spectra of X1. Only the two longest \emph{XMM-Newton}  observations of X1, however, have $\gtrsim$2000-counts spectra, which could be responsible for the non-detection of curvature above 2\,keV. We therefore caution that the interpretation of the MCD plus power-law model fits of X1 presented in \S~\ref{LT} depends on the assumption that the hard band spectrum is well represented by a simple power-law arising in an optically thin corona, while the soft emission arises directly from the accretion disk.

\begin{deluxetable*}{cccccccc}
\tablewidth{0pt}
\tablecaption{NGC\,1365\,X1: Optical and IR Limits}
\tablehead{\colhead{Telescope} &\colhead{Filter} &\colhead{$T\si{exp}$} &\colhead{Aperture} &\colhead{MJD} &\colhead{ULX limit} &\colhead{starE} &\colhead{starW}\\
\colhead{(1)} & \colhead{(2)} & \colhead{(3)} & \colhead{(4)} & \colhead{(5)} & \colhead{(6)} & \colhead{(7)} & \colhead{(8)}  }
\startdata
VLT\,FORS-2	& $R$-special	& 30	& $1.5''$ & 52736 & $R>23.3$	& $18.5\pm0.3$ & $19.4\pm0.3$ \\
HST\,WFPC2	& F814W		& 10	& $0.8''$ & 49732 & $I>21.4$		& $16.81$ & $17.56$ \\
HST\,WFPC2	& F814W	   	& 10	& $0.8''$ & 49732 & $I>21.6$		& $16.96\pm0.03$ & $17.71\pm0.05$ \\
HST\,WFPC2	& F555W		& 100&$0.8''$ & 51220 & $V>24.1$	& $19.8\pm0.1$ & $20.4\pm0.1$ \\
2MASS		& $J$		& ...	& $3.0''$ & 51602 & $J>17.2$	& $15.18\pm0.07$ & $16.32\pm0.17$ \\
2MASS		& $H$		& ...	& $3.0''$ & 51602 & $H>16.2$	& $14.56\pm0.06$ & $15.79\pm0.21$ \\
2MASS		& $K$		& ...	& $3.0''$ & 51602 & $K>15.9$ 	& $14.50\pm0.10$ & $15.32\pm0.20$ \\
\enddata
\tablecomments{ (1) and (2): Telescope/instument configuration and filter used in each observation; (3) Exposure time, $T\si{exp}$, in seconds; (4) Aperture radius used to measure source fluxes; the background was measured locally within an aperture with twice the area of the source aperture; (5) Modified Julian Date (MJD) of the observation; (6) ULX limit in the standard Johnson-Cousins filters; (7) and (8) Magnitudes of the two stars (to the East and West of the ULX position, respectively; see Figure~\ref{x1image}) used for magnitude calibration in each Johnson-Cousins filter.}
\label{opt}
\end{deluxetable*}

\begin{figure*}
\plottwo{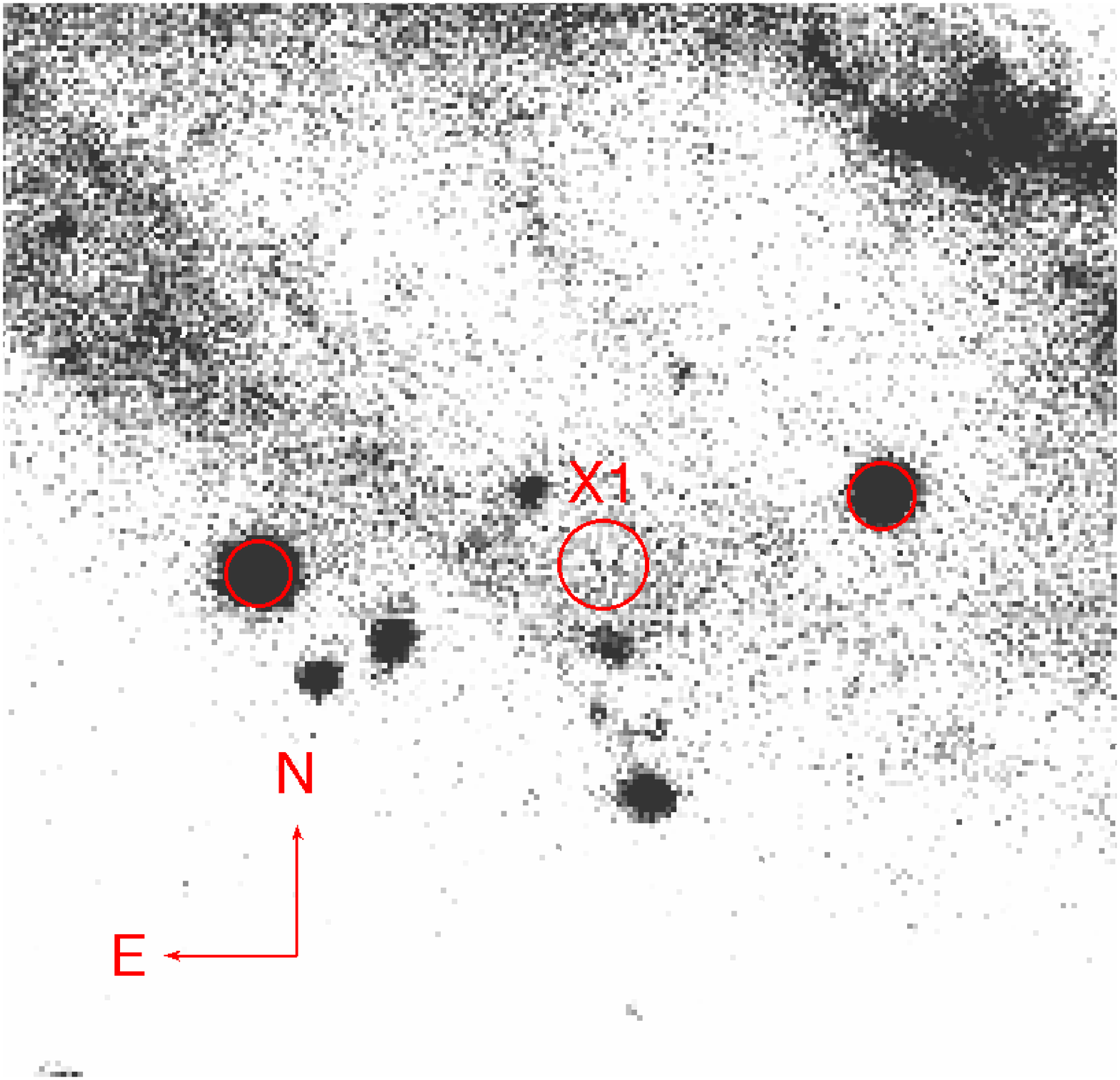}{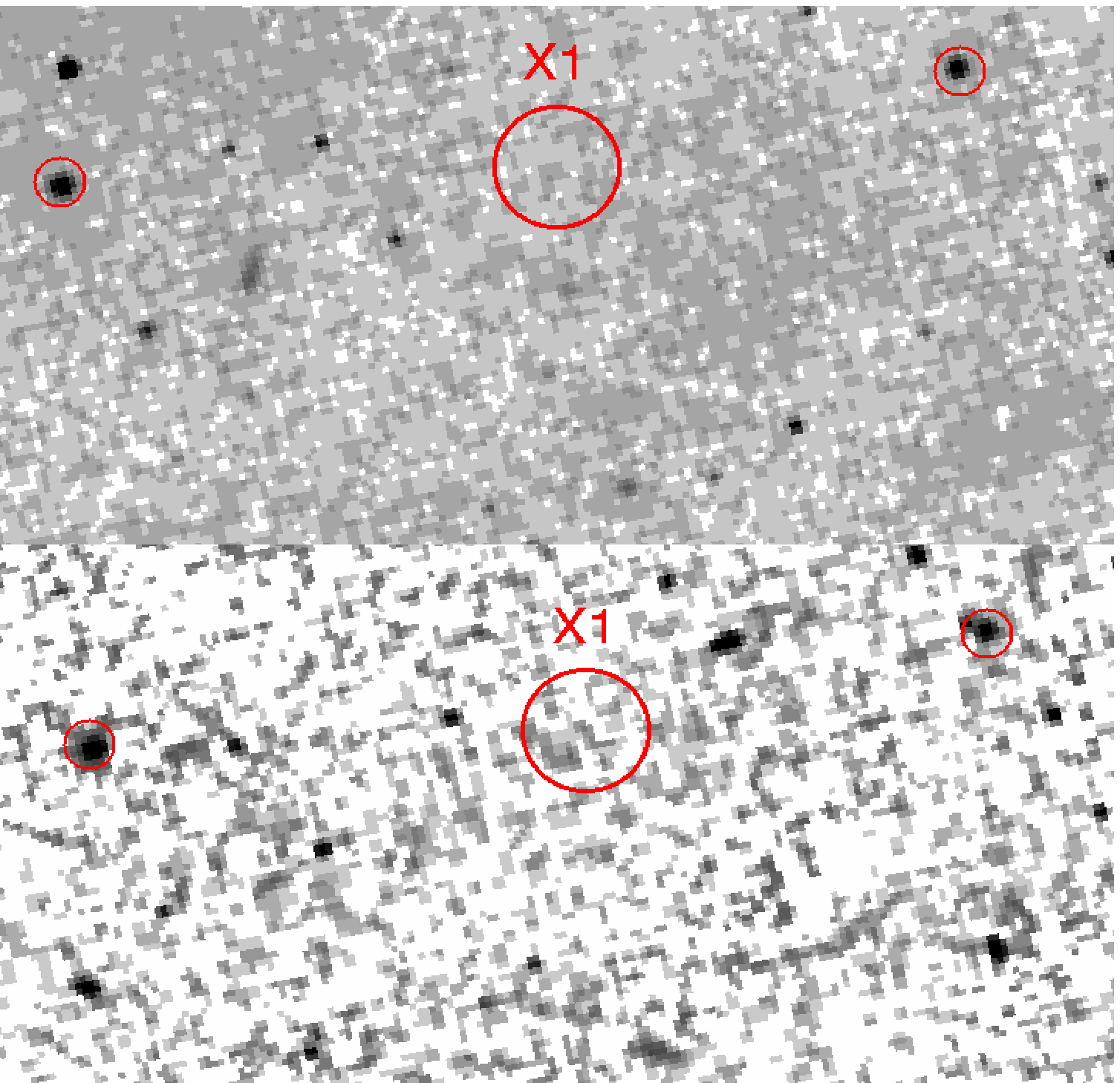}
\caption{The optical region around NGC\,1365\,X1 through the VLT-FORS2 R-band filter (left) and the HST F555 (top right) and F814 (bottom right) filters. The ULX is marked by 2$''$-radius circles on all images; $1.5''$ ($0.8''$) circles denote the W and E stars on the VLT (HST) images (see \S~\ref{x1_optical}). The intensity of the VLT image is scaled to reveal the diffuse extended emission at the position of the ULX as well as the two faint point sources 3.7$''$S and $4''$NE of the X-ray position of X1.
\label{x1image}}
\end{figure*}

\subsection{X1: Optical/IR Observations}
\label{x1_optical}

X1, whose X-ray position is known to better than $1''$ based on the \emph{Chandra} images, was not detected on any of the optical or IR images available. Nevertheless, we use these images to obtain sensitive limits to the $R$, $I$, $J$, $H$, and $K$ Johnson-Cousins band magnitudes of the ULX counterpart. The details of the lower-limit estimates are presented below and summarized in Table~\ref{opt}. We correct all optical and IR images astrometrically using the USNO-B catalog stars \citep{usno} in the immediate vicinity of the ULX source position.

The left panel of Figure~\ref{x1image} shows a 30\,s VLT FORS2 (Very Large Telescope FOcal Reducer and low dispersion Spectrograph) R-Special filter\footnote{The R-special filter is a Johnson-Cousins filter, slightly shortened at the red end to avoid sky emission lines.} observation of NGC\,1365 within 50$''$ of the X-ray position of X1. Smaller region cutouts around the X1 X-ray position in two of the HST filters are shown in the right panel of Figure~\ref{x1image}. The ULX X-ray position coincides with a faint spiral extension; two faint sources are seen $3.7''$S and $4''$NE of the X-ray position, respectively. Considering the large separations, neither of these sources is a likely optical counterpart, but the optical counterpart of the ULX was clearly fainter than both sources during the observation. The two bright stars $\sim16''$E and $\sim13''$W of the ULX in Figure~\ref{x1image} have USNO-B\,1.0 R-band magnitudes of $18.5\pm0.3$ and $19.4\pm0.3$, respectively. Judging by the flux ratios between those stars and the faint sources $\sim4''$ S and NE from the ULX X-ray position, the faint sources have R-band magnitudes of $\sim22.4$ and $\sim22.3$, respectively. We chose a local background region a few arcsec NW of NGC\,1365\,X1 (but still within the diffuse spiral arm emission) and aperture radii of $1.5''$ for all flux measurements. The R-band magnitude limit for the ULX emission is $R>23.3$. 

Two HST\,WFPC2 observations of the central region of NGC\,1365 (see the right panel of Figure~\ref{x1image}) were taken in 1995 with the F814W (10\,s) and F555W filters (100\,s). The HST F814W filter is close to the Johnson-Cousins $I$-band. Using the $I$-magnitudes of the same two bright stars (E and W of the ULX source position in Figure~\ref{x1image}) as photometric calibrators, we estimate a lower limit to the apparent magnitude at the ULX position of $I>21.4$ (see the 2d row of Table~\ref{opt}). If we use the independent HST\,WFPC2 photometric calibration,\footnote{http://www.stsci.edu/documents/dhb/web/c03\_stsdas.fm3.html} we obtain the values given in the 3d row of Table~\ref{opt}, and a ULX source position lower limit of $I>21.6$, in agreement with the lower apparent-magnitude limit obtained using bright-star calibration. The lower limit to the $V$-band apparent magnitude at the ULX  position given in the 4th row of Table~\ref{opt} was estimated in similar fashion using the 100\,s F555W-filter HST exposure.

2MASS coverage of the region around the ULX source position is also available. We use relative photometry (using the same bright stars E and W of the source) to estimate the $H$, $J$, and $K$ band lower limits. The results, using $3''$-radius source aperture measurements, appear in the last three rows of Table~\ref{opt}.  We comment on the implications of all optical/IR apparent magnitude limit results in \S~\ref{Oratios}.

\begin{figure}
\plotone{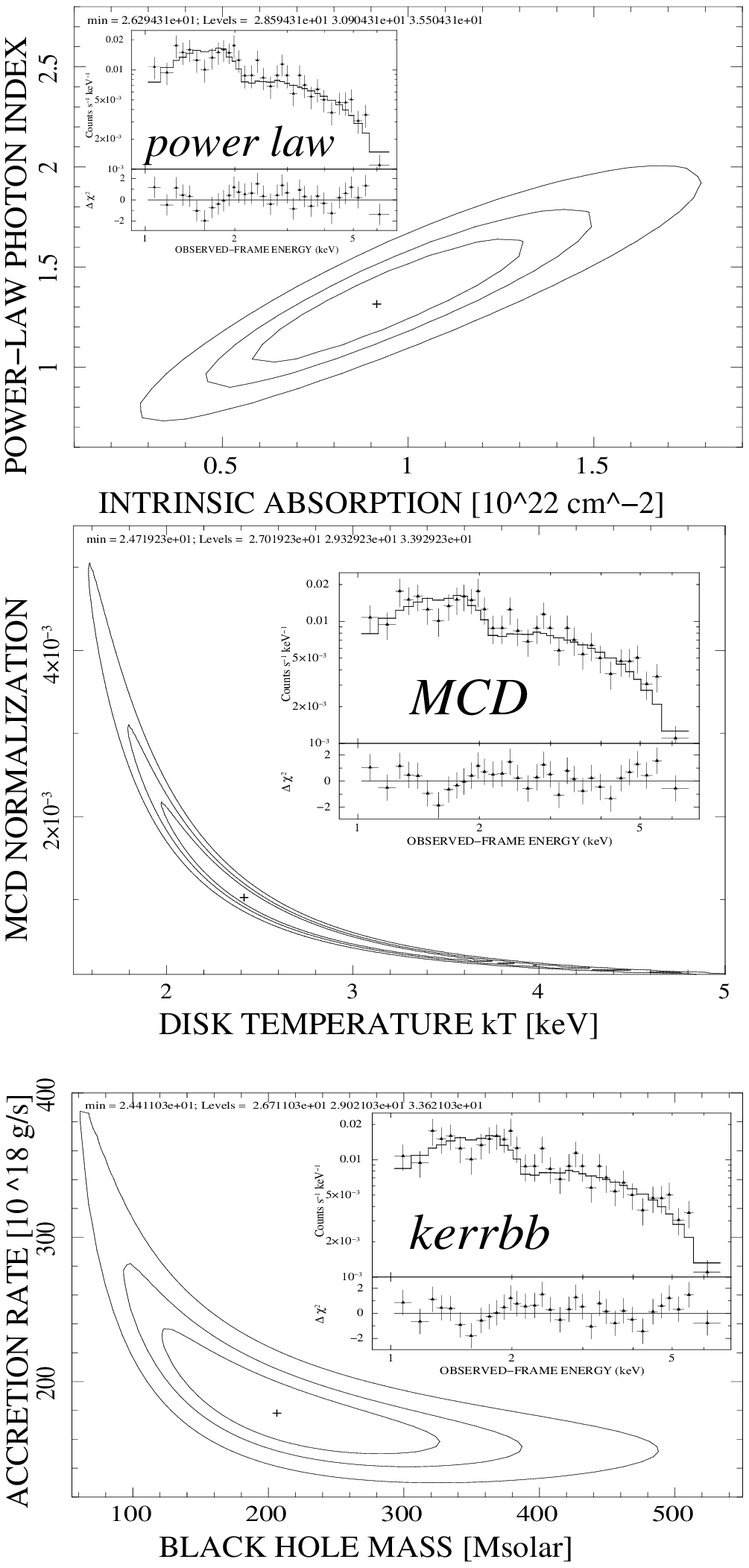}
\caption{NGC\,1365\,X2: X-ray spectroscopic fits to the highest S/N observation, \emph{Chandra} Obs.ID.~6870. All fits include an intrinsic absorber with $N\si{H,i}=6-9\times10^{21}$\,cm$^{-2}$ as well as Galactic absorption (see \S~\ref{x2LT} for details). \emph{Top:} Power-law fit (inset) and  3$\sigma$ confidence contours for the power-law photon index and the column density of the intrinsic absorber. \emph{Middle:} MCD fit (inset) and 3$\sigma$ confidence contours for the normalization and the maximum color temperature. \emph{Bottom:} The \emph{kerrbb} fit (inset) and  3$\sigma$ confidence contours for the black hole mass and accretion rate. The spectral fits, shown as insets, give the data overlaid with the model in the upper panels and the fit residuals in the lower panels.
\label{x2spec}}
\end{figure}

\begin{figure}
\plotone{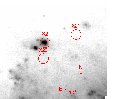}
\caption{The optical region around NGC\,1365\,X2 through the VLT-FORS2 R-band filter. The X-ray point sources are denoted by $3''$-radius circles. X2 coincides with the extended emission of an HII region.}
\label{x2image}
\end{figure}

\section{NGC\,1365\,X2}
\label{x2}

NGC\,1365\,X2 (03:33:41.85, -36:07:31.4) is a highly variable X-ray point source in the north tip of the eastern spiral arm of NGC\,1365 (see Figure~\ref{Ogalaxy}). Based on the highest luminosity reached during one of the April 2006 \emph{Chandra} observations, which exceeds $4\times10^{40}$\,erg\,s$^{-1}$ in the \hbox{0.5--10\,keV} band, the object is among the most luminous ULXs known. 

We extracted spectra for the X2 source from all \emph{Chandra} observations as described in \S~\ref{xspec} for NGC\,1365\,X1. Because of the proximity of X2 to other X-ray point sources (the closest one is only $9''$S of X2, the second closest $20''$W) combined with the larger \emph{XMM-Newton} point spread function and the faintness of X2 during the \emph{XMM-Newton} observations, we do not attempt to extract spectra/upper limits from these exposures. We attempted to fit the \emph{Chandra} spectra with both power-law and simple MCD models. Assuming galactic absorption, we find the simple power-law or MCD fits are unacceptable and intrinsic absorption is necessary in both cases. The absorbed power-law fits are summarized in Table~\ref{x2pl}. Due to the small number of counts, we use the Cash-statistic fit for all but one of the spectral fits -- the highest signal-to-noise (S/N) spectrum (Obs.ID.~6870), which has 535 counts in the 0.3-10\,keV band, was binned to 15 counts per bin and fit using $\chi^2$ statistic.  We show the power-law fit and the 3$\sigma$ confidence contours for the photon index and the column density of the intrinsic absorber for the highest S/N fit of X2 in the top panel of Figure~\ref{x2spec}. For all spectra with less than $\sim 200$ counts in the 0.3--10\,keV band, we fix the power-law spectral index to $\Gamma=1.3$, the value obtained for the two spectra with the largest numbers of counts. A cold absorber of  $N\si{H,i} = 6-9\times10^{21}$\,cm$^{-2}$ is required by all fits, including the MCD fits. 

The absorbed MCD fits are statistically indistinguishable from the absorbed power-law fits presented above. For all April 2006 \emph{Chandra} observations the maximum color temperature estimated from the MCD fits was around 2\,keV, with large uncertainties ($1\sigma$ uncertainties $\sim0.3-0.6$\,keV).  The $3\sigma$ confidence contours for two interesting parameters for the highest S/N MCD fit  are shown in the middle panel of Figure~\ref{x2spec}. The addition of a comptonized component (e.g. a power law) does not improve the fits. We caution that, considering the small number of total counts and even smaller number of counts above 5\,keV, it is conceivable that the absorbed MCD model is inappropriate, and the disk-temperature constraints are unreliable. ULXs and X-ray binaries whose MCD fits result in unusually high disk temperatures similar to those obtained for X2 have been known since \citet{M00}. \citet{M00} suggested that the unusually high temperatures are naturally explained by emission from a disk around a rotating black hole. The bottom panel of Figure~\ref{x2spec} shows a multi-temperature blackbody model fit to the highest S/N X2 spectrum which assumes a thin, steady-state, general relativistic accretion disk around a rotating (Kerr) black hole \citep{Li05}. Both the MCD and the \emph{kerrbb} fits shown in Figure~\ref{x2spec} are acceptable ($\chi^2/DoF=25/31$ and $\chi^2/DoF=24/32$, respectively) and we comment on the results in \S~\ref{x2LT}.

X2 falls within a 1995 \emph{ROSAT} HRI observation. \citet{Singh99} claim a 3.5$\sigma$ detection with $0.0020\pm0.0006$\,cps ($F\si{0.5--10\,keV}\sim1.3\times10^{-12}$\,erg\,s$^{-1}$\,cm$^{-2}$ assuming a power law model with $\Gamma=1.3$ and $N\si{H,i}=9\times10^{21}$\,cm$^{-2}$). We reanalyzed the \emph{ROSAT} HRI observation and find only a very faint source with  $0.00053\pm0.00026$\,cps at that position ($\sim 2 \sigma$); we use this as an upper limit  to the X-ray count rate of X2 at this epoch, arriving at an obscuration corrected flux of $F\si{0.5-10\,keV}<3.5\times10^{-13}$\,erg\,s$^{-1}$\,cm$^{-2}$ for a power law model with $\Gamma=1.3$ and $N\si{H,i}=9\times10^{21}$\,cm$^{-2}$. X2 is one of the three sources that were not detected during the 2002 \emph{Chandra} observation with an upper limit of CR$<0.0004$\,cps or \hbox{$F\si{0.5--10\,keV}<0.9\times10^{-14}$\,erg\,s$^{-1}$\,cm$^{-2}$}.

\section{IMBHs, super-Eddington HMXBs, or Background AGNs?}
\label{models}

Super- or hyper-novae can reach X-ray luminosities of over $10^{40}$\,erg\,s$^{-1}$, but they are expected to be steady sources on short timescales and to decline in luminosity slowly over periods  of months to years. The strong variability of both NGC\,1365\,X1 and NGC\,1365\,X2 over extended periods of time (factors of 10 over years) and their short time flares argue against supernova emission and imply a compact accreting source in both cases. In addition, neither of the two ULXs discussed here has the  thermal emission-line dominated spectrum from shock-heated plasmas which is characteristic of supernovae.  Below we discuss the IMBH, super-Eddington SMBH, and background AGN possibilities for the two NGC\,1365 ULXs, focusing on their spectral characteristics, luminosity variability, and optical/IR-to-X-ray ratios.

\subsection{Background AGNs?}

We argue that both NGC\,1365\,X1 and NGC\,1365\,X2 are unlikely to be background AGNs. Based on their positions -- X1 coincides with one of the diffuse weak spiral arms south of the NGC\,1365 nucleus, while X2 coincides with an HII region at the tip of the main NE spiral arm, as well as the negligible number of background sources as bright as these ULXs expected for a similar-sized random patch of sky, X1's unusual X-ray-to-optical/IR ratios and its small intrinsic  X-ray absorbing column, an IMBH or a supercritically accreting SMBH are more likely in both cases. We cannot rule out that X1, which has no optical counterpart,  is not an optically/IR obscured AGN with little X-ray obscuration with certainty, but the association of NGC\,1365\,X2 with the \ion{H}{2} region in the NE arm of the galaxy makes it extremely unlikely that the source is a background AGN.

\subsubsection{$log{N}-\log{S}$ Argument}
\label{logNlogS}

According to the $\log{N}-\log{S}$ Lockmann Hole results of  \citet{lhole}, we expect $<0.2$ sources per square degree with  $F\si{0.5--10\,keV}>5-6\times10^{-13}$\,ergs\,s$^{-1}$. Consequently, $<2\times10^{-4}$  sources with flux similar or higher than that of the maximum flux of X1 or X2 are expected within $2'$ of the center of NGC\,1365. This, together with the fact that X1 is located in one of the faint spiral-arm extensions of NGC\,1365 and X2 coincides with an \ion{H}{2} region at the tip of the NE arm, make it unlikely that either of the sources is a background AGN.

\subsubsection{X-ray-to-Optical/IR Ratios} 
\label{Oratios}

\hspace{2.8cm}{\emph{NGC\,1365\,X1}}

As noted in \S~\ref{x1_optical}, any optical counterpart of the X1 source is fainter than $R>23.3$. This limit is consistent with any stellar mass companion, including the brightest supergiants. Computing the X-ray-to-optical ratio, $\log{\textrm{[X/O]}}=\log{F\si{X}}+0.4R+5.61$, we find lower limits between 1.5 and 2.7 for the full range of X-ray fluxes observed for X1. Typical X-ray-to-optical ratios for AGNs range between $-1<\log{\textrm{[X/O]}}<1$ \citep[e.g.,][]{maccacaro88,horn01}, with the upper range corresponding to blazars. About 90\% of all AGNs have $\log{\textrm{[X/O]}}<2$. Since the observed lower limit to the X-ray-to-optical ratio for X1 is  smaller than this value for only 3 of the 12 X-ray observations, we consider it highly unlikely that this ULX is a background AGN. For comparison, the 8 ULXs which were found to be background AGNs (or a narrow line galaxy in one case) by \citet{ulx8qso} have $0.04<\log{\textrm{[X/O]}}<0.6$. The two additional X-ray point sources from Table~\ref{tabSrc} with optical counterparts in the USNO-B\,1.0 catalog have $\log{\textrm{[X/O]}}\sim-0.4$ (X17) and $\log{\textrm{[X/O]}}\sim0.3$ (X15). X15, whose optical counterpart is a point-source, is likely a background AGN.

We estimated a limit of $I>21.6$ ($F_I<6$\,$\mu$Jy at 0.9\,$\mu$m) for the $I$-band magnitude of NGC\,1365\,X1. The X-ray-to-I-band ratio,  $\log[\textrm{X}/I]=\log{[F\si{0.5--2\,keV}/F_I]}$, has a lower limit ranging between $-0.1$ and 1.7 for the full range of observed 0.5--2\,keV fluxes, with a most likely lower limit between 0.4 and 0.7, indicating that the source could be a very rare optically faint/X-ray bright background AGN at a redshift $>$2 \citep[see Figure~4 of][]{BH05}. If that was the case, however, the measured flux would imply an X-ray luminosity of $\sim10^{46}$\,erg\,s$^{-1}$. According to the X-ray luminosity function of \citet{LF} type 1 (unobscured) AGNs with such high luminosities are rare: $\sim10^{-9}$\,Mpc$^{-3}$. The analysis of the X-ray background \citep[e.g.,][]{XRB} further shows that the ratio of obscured to unobscured AGNs is $\lesssim1$ for luminous AGNs, implying that there are $\lesssim200$ AGNs in total of comparable luminosities between redshifts of 1.9 and 2.1 in the whole universe, which together with the large variability  of X1, uncharacteristic for luminous AGNs, make it very unlikely that X1 is a luminous redshift $>$2 AGN.

\begin{figure*}
\plottwo{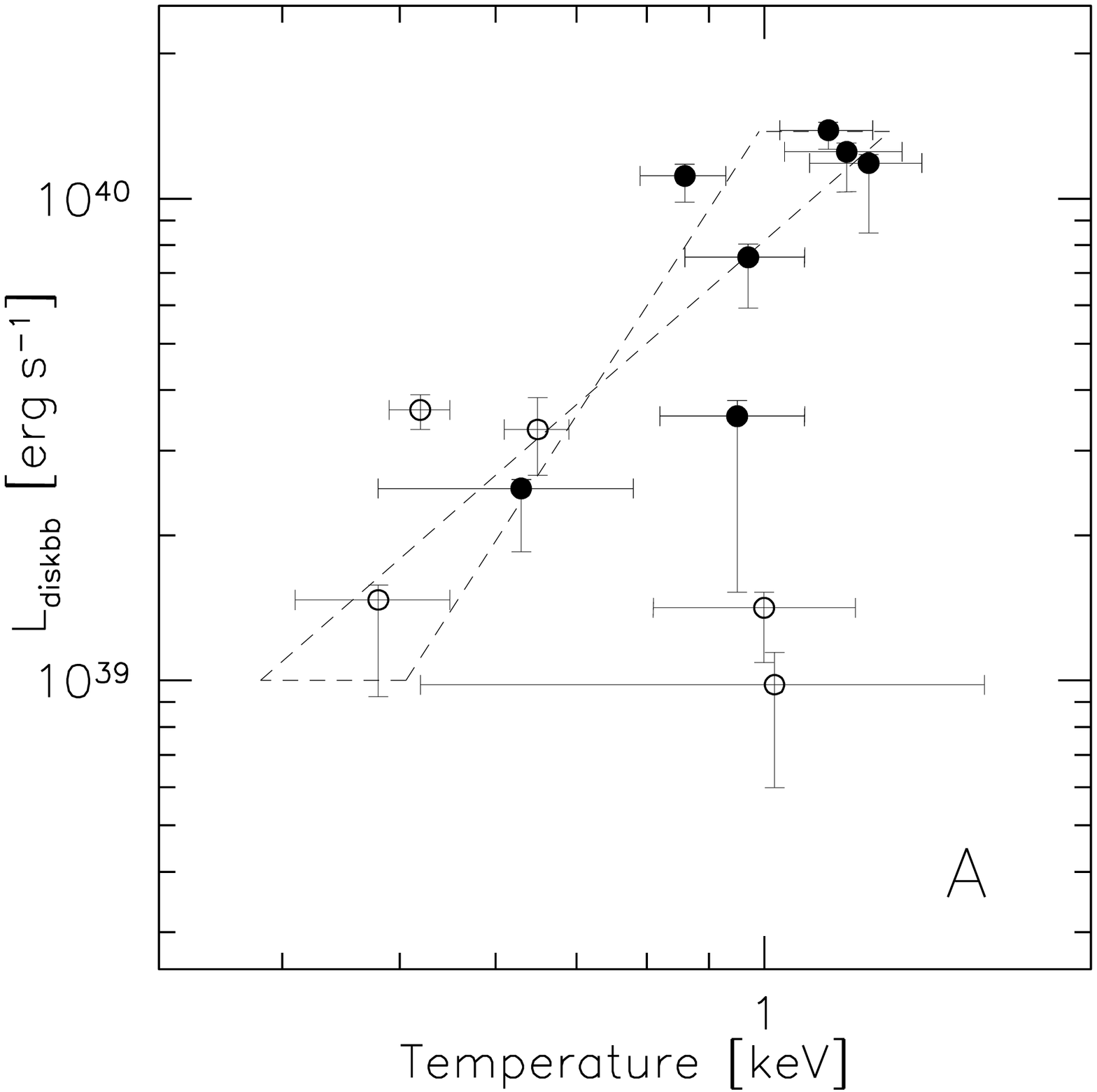}{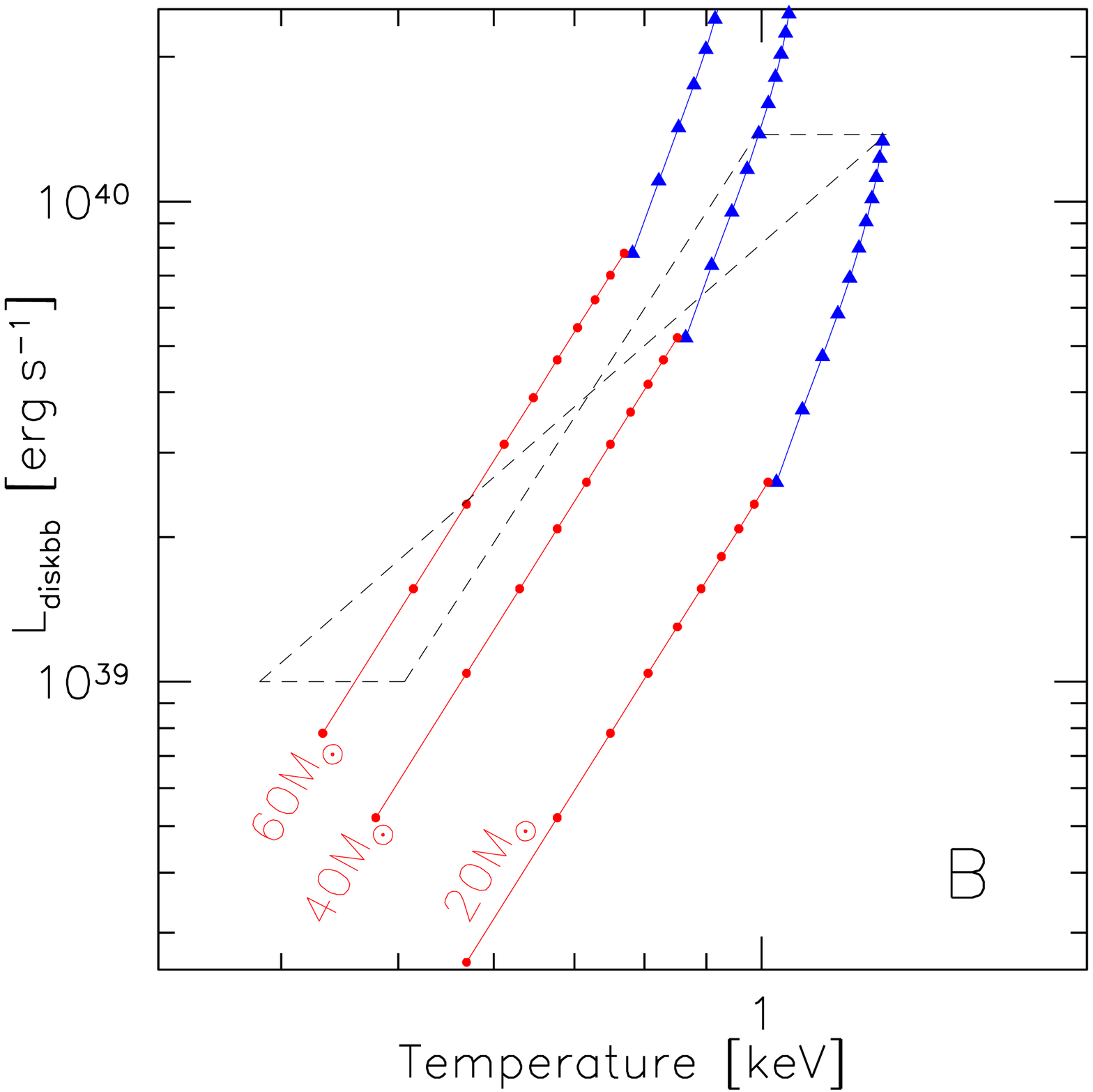}
\caption{NGC\,1365\,X1 -- the multi-color disk (MCD) temperature vs. the accretion disk luminosity: data (\emph{Panel A}) and theoretical models (\emph{Panel B}). The \emph{XMM-Newton}/\emph{Chandra} points are denoted by open/solid circles. A power-law fit to the estimated MCD parameters, $L\propto T^{\alpha}$,  with $\alpha=2.8^{+1.1}_{-0.6}$, is given with the dashed lines in both panels. The blue lines show the (color corrected) maximum temperature ($T\si{max}$) of supercritically accreting black holes with M=20\,M$_{\odot}$,  M=40\,M$_{\odot}$, and M=60\,M$_{\odot}$ \citep{P07}. The accretion rate varies between  $1<\textrm{\.m}<800$ for $T\si{max}$ (each triangle corresponds to a doubling of the accretion rate). The red lines show the $L \propto T^4$ relations expected for 20\,M$_{\odot}$, 40\,M$_{\odot}$, and 60\,M$_{\odot}$ black holes with accretion rates $0.1<\textrm{\.m}<1$ (each dot corresponds to an increase in the accretion rate by 0.1). 
\label{x1LT}}
\end{figure*}

Using the 2MASS observations, the $K\si{S}>15.9$ limit of X1 corresponds to \hbox{$F_{K\si{S}}<0.29$\,mJy} at 2.2\,$\mu$m, indicating a lower limit of the  X-ray-to-$K\si{S}$ flux ratio of $\sim$0.2 (ranging between 0.04 and 0.6 for the observed X-ray range).  \citet{Wilkes02} studied the X-ray properties of red 2MASS AGNs. According to their Figure~1, typical red 2MASS AGNs have X-ray-to-$K\si{S}$  flux ratios $<0.05$, while the $z<1$ broad-line AGNs form \citet{Elvis94} have X-ray-to-$K\si{S}$  flux ratios between 0.05 and 0.7. The lower limits to the  X-ray-to-$K\si{S}$ flux ratio for NGC\,1365\,X1 is consistent with an unobscured broad-line AGN, but this interpretation is inconsistent with the observed optical limits.

\hspace{2.8cm}{\emph{NGC\,1365\,X2}}

In the optical image, X2 is associated with a bright extended emission region in the eastern spiral arm of NGC\,1365 (see Figure~\ref{x2image}), $86''$ S from the NGC\,1365 nucleus (equivalent to $\approx8.8$\,kpc), which was classified as an \ion{H}{2} region by \citet{Hodge}. Using the USNO-B\,1.0 catalog's $R$-band magnitudes of the optical counterpart, $R1=13.25$ and $R2=13.75$, we estimate an X-ray-to-optical ratio of $-2<\log{\textrm{[X/O]}}<-1$, depending on the X-ray flux level. The \ion{H}{2} region has been studied in detail by \citet{HII} and \citet{swara}. It is a typical large star-forming region with a diameter of 670\,pc, $M\si{V} =-12.4$, and an H$\alpha$ line luminosity of $1.6\times10^{39}$\,erg\,s$^{-1}$ (S. Ravindranath, private communication). \citet{HII} study the optical spectrum of the \ion{H}{2} region (number 33 in their Table~2) in comparison with other \ion{H}{2} regions in NGC\,1365. The observed narrow-line ratios suggest an ionization-bound nebula which is photoionized by massive stars. There is no evidence that the X-ray emission of NGC\,1365\,X2 affects the observed line ratios, which is consistent with its interpretation as an X-ray binary (whose energy input would be too small to affect the observed narrow line-ratios of a giant nebula). 

According to the X-ray observations, X2 contributes about $5\times10^{49}$ ionizing photons per second, and the H$\alpha$ luminosity emitted as a consequence of this ionizing flux should be about  $5\times10^{37}$\,erg\,s$^{-1}$ \citep[see Section 5.3 of][and references therein for the details of the computation]{SC05}. This is a factor of 30 fainter than the measured  H$\alpha$ line luminosity of the \ion{H}{2} region, confirming the conclusion that X2 does not contribute substantially to the photoionization of the surrounding nebula.

\subsection{$L$-$T$ Diagram Results}
\label{LT}

\subsubsection{NGC\,1365\,X1}
Panel A of Figure~\ref{x1LT} shows the Luminosity-Temperature ($L$-$T$) diagram for NGC\,1365\,X1. The estimated MCD temperatures and accretion disk-luminosities follow a $L \propto T^{\alpha}$ relation,  with $\alpha=2.8^{+1.1}_{-0.6}$.  Panel B of Figure~\ref{x1LT} displays the theoretical tracks for 20, 40 and 60\,M$_{\odot}$ black hole primaries, accreting at sub-Eddington (colored red) or super-Eddington rates \citep[colored blue;][]{P07}. The X1 relation agrees with the $L \propto T^4$ relation expected for a black hole primary with mass between 40\,M$_{\odot}$ and 60\,M$_{\odot}$, but the observed maximum luminosities during the \emph{Chandra} flare of April 2006 ($L\si{X}>10^{40}$\,erg\,s$^{-1}$) require super-Eddington accretion rates and/or beaming.
In the \citet{P07} model, the observed luminosities can reach \hbox{$L \approx L\si{Edd}(1+0.6\ln{\textrm{\.m}})$} where \.m is the accretion rate in units of Eddington.\footnote{In the supercritical accretion model of \citet{Ohsuga}, the luminosity can exceed the Eddngton luminosity by a factor of up to $\sim$10 for face on view and \.m=100--1000.} The maximum color temperature of the disk is \hbox{$T\si{c,max}=f\si{c}\times1.6\times(m^{-1/4})\times(1-0.2\textrm{\.m}^{-1/3})$\,keV}, where $m$ is the black hole mass in units of M$_{\odot}$, and f\si{c} is the color-correction factor, $f\si{c}=1.7$ \citep[$T\si{max}$ is the maximum temperature of the advective disk, see eqn.~36 of][]{P07}.  A 40--60\,M$_{\odot}$ black hole primary is consistent with all \emph{XMM-Newton} and \emph{Chandra} points in Figure~\ref{x1LT} within $3\sigma$ of their estimated uncertainties if the accretion rate varies between 0.1 and 16 of the Eddington rate in the \citet{P07} model. The measured temperatures during the \emph{Chandra} flare, are systematically high relative to the super-Eddington 60\,M$_{\odot}$ model (by about 30\% or $\sim3\sigma$). A variation of the color-correction factor with accretion rate for 1$<$\.m$<$16 could account for this discrepancy. 

We note that a less massive primary is problematic not only because it fails to explain the observed temperature-luminosity pairs in Figure~\ref{x1LT}, which could, as noted in \S~\ref{xspec}, be affected by the specific model fit assumptions. A 10\,M$_{\odot}$ black hole primary, for example, is also problematic because the required beaming factors or super-Eddington accretion rates (a beaming of over a factor of 30 or, alternatively, $>10^{-6}$\,M$_{\odot}$ available for accretion in $\sim$2\,days during the \emph{Chandra} flare in the Poutanen et al. 2007 model) are uncomfortably large. 

We conclude that the $L$-$T$ diagram of X1 is consistent with the expectations for a \hbox{40--60\,M$_{\odot}$} black hole primary accreting at 0.1$<$\.m$<$16 from a massive companion star, offering no support for the IMBH hypothesis.

\subsubsection{NGC\,1365\,X2}
\label{x2LT}

As described in \S~\ref{x2}, the X2 spectra are consistent with absorbed power-law, or absorbed MCD fits, and the current data is of insufficient quality to distinguish between these or more complex models (e.g. models including a combination of a MCD and a comptonized component).
The MCD fits to the X2 spectra are inadequate to constrain the the maximum color temperature and inner radius of emission for each individual observation and assuming a simple MCD model could lead to unreliable temperature estimates. As a result, the following interpretation of the results depends critically on the assumption that X2 is in a disk-dominated state and the MCD or \emph{kerrbb} models are appropriate.

\begin{deluxetable*}{ccccccc}
\tablecaption{NGC\,1365\,X2: Absorbed Power-Law Fits}
\tablehead{\colhead{ObsID} &\colhead{$F\si{2--10\,keV}$} &\colhead{$F\si{0.5--2\,keV}$} &\colhead{$\Gamma$} &\colhead{N$_{\textrm{\scriptsize{H}}}$} &\colhead{Data/Method} &\colhead{stat/DoF}\\
\colhead{(1)} & \colhead{(2)} & \colhead{(3)} & \colhead{(4)} & \colhead{(5)} & \colhead{(6)} & \colhead{(7)} }
\startdata
3554  & $<0.7$  & $<0.2$ & 1.3 fixed    & 9 fixed & ACIS-S    & ...   \\
6871  & $5.6_{-3.3}^{+3.8}$ & $1.7_{-0.6}^{+0.6}$ & 1.3 fixed & $7\pm3$ & ACIS-S/C-stat    & 40/44  \\
6872  & $42_{-29}^{+9}$ & $12_{-9}^{+4}$ & $1.3\pm0.2$ & $9\pm2$ & ACIS-S/C-stat & 165/228 \\
6873	  & $37_{-8}^{+8}$ & $11_{-1.6}^{+2.0}$ & 1.3 fixed & $10\pm1$ & ACIS-S/C-stat & 175/187 \\
6868  & $25_{-8}^{+7}$ & $7.4_{-1.9}^{+1.9}$ & 1.3 fixed& $9\pm2$ & ACIS-S/C-stat &122/151 \\
6869  & $12_{-4}^{+5}$ & $3.6_{-1.2}^{+1.0}$ & 1.3 fixed & $8\pm1$ & ACIS-S/C-stat  & 73/102\\
6870  & $63_{-15}^{+10 }$ & $20_{-12.4}^{+4.4}$ & $1.3\pm0.2$  & $9\pm2$    & ACIS-S/$\chi^2$ & 26/31\\
\enddata
\tablecomments{The fluxes are quoted in units of $10^{-14}$\,erg\,s$^{-1}$\,cm$^{-2}$. (1) Observation ID; (2) Absorption-corrected flux in the 2--10\,keV band; (3) Absorption-corrected flux in the 0.5--2\,keV band; (4) Power-law photon index;  (5)  Intrinsic absorption column density in units of $10^{20}$\,cm$^{-2}$; (6) Data and fitting method used; (7) $\chi^2/DoF$ or C-stat$/DoF$ of the fit.}
\label{x2pl}
\end{deluxetable*}

Assuming that X2 is in a disk dominated state and the MCD model is appropriate, the disk temperature estimates are too high for a sub-Eddington accretion around a non-rotating black hole (\hbox{$T\si{c,max}<1.7$\,keV}). The \citet{P07} model with a SMBH primary can reproduce the observed disk temperatures for $\textrm{M}<5$\,M$_{\odot}$, but  cannot reproduce the maximum luminosity (a factor of $\gtrsim$50 higher), since in that model the luminosity exceeds Eddington by a factor of $1+0.6\ln{\textrm{\.m}}$, which is less than a factor of 5 for $\textrm{\.m}<10^3$. Other supercritical accretion models \citep[e.g.][]{Ohsuga} also predict luminosities which exceed the Eddington by factors of a few, up to 10--15.  If we would believe the high accretion disk temperatures, \citet{M00} have suggested a solution -- a rotating black hole would naturally explain the higher disk temperatures. For X2 a maximally rotating black hole ($a=0.998$) with disk inclination fixed to $\theta=85$ (close to edge on), we obtain $\textrm{M}=210\pm70\textrm{\,M}_{\odot}$ with an accretion rate \hbox{$\textrm{\.M}=(1.8\pm0.3)\times10^{20}$\,g\,s$^{-1}$} ($\textrm{\.m}\approx0.4-0.5$), and an intrinsic absorption of $N\si{H,i}=6\pm2\times10^{21}$\,cm$^{-2}$. The bolometric luminosity of this model is $\sim1-2\times10^{40}$\,erg\,s$^{-1}$, similar to the \emph{kerrbb}-fit value of \hbox{$3.5^{+0.4}_{-1.0}\times10^{40}$\,erg\,s$^{-1}$}. 
Consequently the X2 spectra are consistent with emission from a disk around a maximally rotating black hole, for a large range of IMBH masses, $\textrm{M}\sim80-500\textrm{\,M}_{\odot}$. The S/N however is not sufficient to distinguish those models from simple absorbed power-laws, and the evidence provided by this spectroscopic fit is only suggestive.

\begin{figure}
\plotone{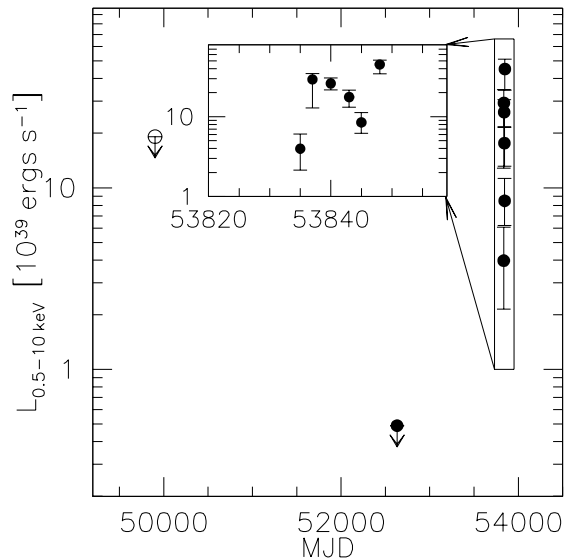}
\caption{NGC\,1365\,X2 luminosity variation with time. The \emph{Chandra} data is shown with solid circles, the \emph{ROSAT}-HRI limit with an open circle; the arrows indicate upper limits.
\label{x2lum}}
\end{figure}

\subsection{Implications of the X-ray Variability}

The X-ray spectra of classical X-ray binaries initially get softer as their flux increases, which is not observed  for either NGC\,1365\,X1 or X2. If anything, the X-ray spectrum of NGC\,1365\,X1 gets harder in the higher flux states. Other ULXs are known to show this behavior on shorter timescales which was suggested to arise in cool-accretion disk plus optically thick corona models of ULX emission as the corona is heated in the high flux states \citep[e.g. NGC\,5204\,X1;][]{Roberts06}. \citet{Soria} interpret the hardening with increased flux as the effect of an additional ionized-absorption component -- a flux decrease in the 0.5--1.5\,keV region which looks like a continuum hardening in power-law fits with neutral absorption.

From Figure~\ref{x1lum}, NGC\,1365\,X1 appears to be persistently luminous in the X-ray band (typically a few times $10^{39}$\,erg\,s$^{-1}$) over decades with occasional short ($\sim$days), but not very powerful ($\lesssim$ a factor of 10) flares. NGC\,1365\,X2, on the other hand, has more transient behavior -- it was not detected before April 2006, when it suddenly reached unobscured  luminosities of a few times $10^{40}$\,erg\,s$^{-1}$, varying on a timescale of days (Figure~\ref{x2lum}). Assuming  either a SMBH or an IMBH binary, the persistent high state of X1 can be interpreted as evidence of a massive donor star ($\gtrsim$20--30M$_{\odot}$) supplying a large, stationary, but not necessarily steady (considering the \emph{Chandra} flare on timescale of days) accretion disk. As noted in \S~\ref{LT} above, the 2006 \emph{Chandra} flare of X1 is too small to be associated with a disk instability, as observed in SXTs in LMXBs, which result in accretion rate and luminosity increases of factors of $\sim10^4$.

\citet{Kalogera} studied the long term transient behavior of SMBHs and IMBHs due to thermal-viscous disk instability for ULXs in young stellar populations.  For the most likely donors in such environments, with masses in excess of 5\,M$_{\odot}$, they inferred a minimum black hole mass for transient behavior, which is $\gtrsim50$\,M$_{\odot}$. Based on this argument and the observed variability patterns of the two ULXs, it is more likely that NGC\,1365\,X2, which is a transient source associated with an \ion{H}{2} region, contains an IMBH.

\subsection{Free-Floating Supermassive Black Hole?}

Recent numerical relativity simulations of  back hole mergers suggest that large gravitational recoil velocities of up to several 1000\,km\,s$^{-1}$ are possible for  mergers of maximally spinning, equal-mass, black holes with anti-aligned spins in the orbital plane \citep{koppitz07, campanelli07a, gonzalesetal07, herrmann07, baker07}. Alternatively, a gravitational slingshot of three or more supermassive black holes can result in the ejection of one of them from the nucleus \citep{Hut92,HL07}. Luminous off-nuclear X-ray sources are potential counterparts to these recoiling black holes \citep[e.g.,][]{Madau04,libeskind06}. Is it possible that either X1 or X2 is a ``free-floating" supermassive black hole which was expelled from the center following a merger with a large recoil velocity or a gravitational slingshot in the past?

A recoiling BH with a kick velocity below the escape velocity will oscillate within the galaxy with a spatial amplitude that depends on the kick velocity \citep{Madau04, gualandris08}. Oscillations typically damp within a time scale of $\sim10^{6-8}$ years, which is comparable to the time scale of a galaxy merger. At present, we do not see evidence for the occurence of a recent merger in NGC\,1365. In particular, a nearly equal-mass merger would have resulted in the formation of a large bulge and randomized the central velocities, contrary to the observed strong bar pattern in NGC\,1365.

Furthermore, within the recoil scenario, we would \emph{not} expect a central black hole in NGC\,1365 at the current epoch. However, NGC\,1365 does harbour an accreting supermassive black hole at its core, as is evidenced by the presence of broad Balmer lines and AGN-typical emission-line ratios \citep[e.g.,][]{schul99} and by the detection of rapid X-ray variability \citep{R07}. In addition, the coincidence of X2 with an \ion{H}{2} region would be hard to explain under the recoil/slingshot-ejection scenarios. With respect to X1 and any other off-nuclear point source not associated with a dense region, assuming that its black hole mass does not exceed \hbox{10$^{5}$\,M$_{\odot}$}, additional evidence against the free-floating black-hole interpretation comes from the availability of accretion material. To reproduce the observed luminosities with a $10^5$\,M$_{\odot}$ black hole (assuming a temperature of $10^4$\,K) we would need to embed it in a medium with density of at least $n\approx1$\,cm$^{-3}$, which is typical for a normal HII region, not the diffuse spiral arm where X1 is located. For the typical interstellar medium which has higher temperatures and lower densities ($n\sim10^{-3}$\,cm$^{-3}$ and $T\sim10^6$\,K), the Bondi accretion rate is a factor of $10^{6}$ lower. While ejected black holes can be temporarily fueled by stellar tidal disruptions or stellar mass loss from the stars which are bound to them upon recoil \citep{KM08}, these events are rare and are very unlikely to operate in two sources simultaneously in one nearby galaxy. Based on all of these arguments, none of the off-nuclear X-ray sources of NGC\,1365 appears to be a compelling recoil/slingshot-ejection candidate. It is still possible, however, that recoils and/or black holes ejected by a gravitational slingshot hide among the ULX population as a whole.

\section{Summary and Conclusions}
\label{conclusion}

We present 26 X-ray point sources detected in a series of \emph{Chandra} observations of the giant spiral galaxy NGC\,1365. The cumulative distribution of sources brighter than a given X-ray flux is  compatible with predictions based on the star-formation rate of the galaxy, as expected if high-mass X-ray binaries dominate the extended X-ray emission of star-forming galaxies \citep[e.g.,][]{G04,GG}. 

Two of the X-ray point sources, NGC\,1365\,X1 and NGC\,1365\,X2, are highly ultraluminous, with peak luminosities exceeding 3--5$\times10^{40}$\,erg\,s$^{-1}$. Both ULXs are most likely accreting compact objects (with SMBH or IMBH primaries), since both the super-/hyper- nova origin and the background or ``free floating"  AGN interpretations appear highly unlikely. Both sources are among the most luminous ULXs known and as such, among the most likely candidates for IMBH binaries. Their luminosities and optical counterparts (a lower limit for X1 and a giant \ion{H}{2} region for X2) are consistent with an accreting IMBH in both cases; assuming no beaming and accretion rate equal to the Eddington rate (computed from the highest \emph{Chandra} luminosities), the black hole masses are $\sim250$\,M$_{\odot}$ in the case of X1 and $\sim340$\,M$_{\odot}$ in the case of X2. 

\citet{RMreview} define three outburst states of SMBHs, based on data of galactic BH binaries: (1) the \emph{thermal} or accretion-disk dominated state (with disk fraction of $>80$\%; previously known as the High/Soft state), (2) the \emph{hard} ($1.4<\Gamma<2.1$) or power-law dominated state (the Low/Hard state), and (3) the \emph{steep power law} (SPL) state, in which both disk emission (typically $<$50\%, but up to 80\%) and a steep power law ($\Gamma>2.4$) contribute (the Very high state). Active nuclei with supermassive black holes are typically observed in the hard state, in which case a radio jet is often present in the case of X-ray binaries as well as low-luminosity AGNs. It is not clear whether accreting IMBHs will exhibit the three states observed in galactic BH binaries.

The observed X-ray spectra for  X1 and X2 are consistent with power-law models including intrinsic absorption, as well as thermal accretion-disk (plus a power law in the case of X1, which contributes about 50\% of the flux) emission models. The 12 \emph{Chandra} and \emph{XMM-Newton} spectra of X1 are equally well fit by power-law emission modified by intrinsic neutral absorption and a multi-color disk (MCD) emission plus about equal contribution from a comptonizing corona (reminiscent of the SPL/Very-high state). The MCD fits to the X-ray spectra of X1 imply maximum color temperatures which are too high for the black hole masses inferred from the maximum black-body luminosity ($\sim100$\,M$_{\odot}$), if we assume isotropic emission and sub-Eddington accretion rates, suggesting that beaming and super-critical accretion also play a role. Assuming that the MCD plus power-law fits provide an accurate representation of the spectra, a black-hole mass primary in the 40--60\,M$_{\odot}$ range is consistent with the observed $L$-$T$ diagram for X1. 

In the case of X2, the observed X-ray spectra are consistent with intrinsically absorbed power-law, MCD, and maximally-rotating IMBH fits and the insufficient photon statistics prevents us from obtaining reliable accretion-disk temperature estimates. The MCD temperature estimates obtained ($\sim$2\,keV) are unrealistically high for even the super-critically accreting SMBH models, suggesting that the simple power-law model or the rotating black hole model provide more viable explanations for the observed X-ray spectra. A rapidly-rotating ($a\sim0.998$) IMBH with $\textrm{M}=$80--500\,M$_{\odot}$ is consistent with the highest S/N \emph{Chandra} spectrum of X2 and can reproduce the observed luminosities. An IMBH interpretation for X2 is also consistent with the theoretical expectations of \citet{Kalogera}, who inferred a minimum black hole mass for transient behavior of $\textrm{M}>50$\,M$_{\odot}$ for a binary with a donor star more massive than $5$\,M$_{\odot}$.

Absorbed power-law models provide equally good fits to the X-ray spectra of X1 and X2. The results of the power-law model fits for X1 and X2 are reminiscent of the hard (formally, Low/Hard) state of galactic BH binaries. The power-law slopes deduced for X1 ($1.6<\Gamma<2.2$) and X2 ($\Gamma=1.3\pm0.2$) are consistent with those of galactic BH binaries in the hard state, where the comptonized component dominates; the observed luminosities are however very high, suggesting that super-Eddington accretion modes and or beaming/collimation also play a role. The maximum luminosities reached (3--5$\times10^{40}$\,erg\,s$^{-1}$) are factors of 10--20 higher than those of the $\lesssim20$\,M$_{\odot}$ black-hole X-ray binaries studied in our Galaxy, pushing the limits of the current supercritical accretion models and requiring very high beaming/collimation.

The most likely black-hole masses are probably around 40--60\,M$_{\odot}$ in the case of X1, and 80--500\,M$_{\odot}$ in the case of X2, although most of the arguments for such high masses come from the thermal components of the spectral fits, which could be unreliable. We consider masses of $\lesssim20$\,M$_{\odot}$ less likely also because they are harder to reconcile with typical expectations for the supercritical/beamed emission models available. 

While this paper was in preparation, \citet{Soria} published an analysis of NGC\,1365\,X1 and concluded that the spectra are best fit by absorbed power-law components in all cases, and an accretion disk component contributes at most 10-25\% of the emission of X1 in select higher S/N spectra. The authors suggest that this is compatible with a $\approx 50-150$\,M$_{\odot}$ mass black hole, where the accretion disk is colder (0.2--0.4\,keV) and the inner parts of the accretion flow are in the form of an outflow or a Comptonizing corona, which is responsible for the power-law emission. The current data, despite its high photon statistics, is insufficient to distinguish between different scenarios without the assumption of a specific theoretical model (e.g., power-law emission with or without contribution from the accretion disk  and intrinsic absorption) followed by a targeted interpretation of the results. Higher S/N X-ray spectra and better constraints on the short term variability of X1, as well as better theoretical understanding of the most luminous ULXs (which will allow us to select the proper theoretical models to fit), will allow us to distinguish those scenarios in the future. It is reassuring  that despite our different interpretations of the X-ray spectra of X1, we reach similar conclusions about the possible black hole mass of X1.

The optical counterpart of X2, a luminous HII region in the spiral arm of NGC\,1365, is the most likely place to find an IMBH -- both in terms of formation mechanisms for IMBHs, which require very dense environments, and in terms of the availability and capture of a suitable donor star to supply the accretion material. The lack of an optical counterpart of X1 makes the association with an IMBH more problematic, but not implausible; recent theoretical work suggests that for primordial IMBHs (one of three possible formation mechanisms, all of which require a dense environment), most IMBHs with masses $<1000$\,M$_{\odot}$ would be ejected from the dense globular cluster where they formed \citep{eject}. Consequently, the lack of an optical counterpart for X1 does not automatically preclude an IMBH primary. The X-ray variability of X2, which appears to be a transient source, is also statistically consistent with an IMBH interpretation. 

\vskip 0.5in \leftline{Acknowledgements}

We wish to thank Hans Ritter and Andrea Merloni for illuminating discussions of the properties of X-ray binaries. We are grateful to Swara Ravindranath for sharing with us her data of the \ion{H}{2} region whose optical position coincides with NGC\,1365 X2 and suggesting we consult \citet{HII} for the optical spectrum. We thank Li-Xin Li for his help with the implementation of the \emph{kerrbb} model. We thank our referee, Roberto Soria, for his careful reading of the manuscript and insightful comments.
 
This work is based on observations obtained with XMM-Newton, an ESA science mission with instruments and contributions directly funded by ESA Member States and the US (NASA). In Germany, the XMM-Newton project is supported by the Bundesministerium f\"ur Wirtschaft und Technologie/Deutsches Zentrum f\"ur Luft- und Raumfahrt (BMWI/DLR, FKZ 50 OX 0001) and the Max-Planck Society. It is also based on an observation obtained with the \emph{Chandra} X-ray telescope (a NASA mission).




\end{document}